\documentclass{article}
\usepackage[latin9]{inputenc}
\usepackage{float}
\usepackage{amsmath}
\usepackage{amssymb}
\usepackage{graphicx}

\makeatletter
\makeatletter
 \@addtoreset{equation}{section}
 
\makeatother

\makeatother

\begin{document}
\begin{titlepage}

\renewcommand{\thefootnote}{\fnsymbol{footnote}}

\begin{flushright}
\begin{tabular}{l}
UTHEP-663
\end{tabular}
\end{flushright}

\bigskip
\begin{center} {\Large \bf  Comments on Takahashi-Tanimoto's scalar solution \\} \end{center}
\bigskip
\begin{center} 
{\large  Nobuyuki Ishibashi}\footnote{e-mail:         ishibash@het.ph.tsukuba.ac.jp} \end{center}
\begin{center} {\it Graduate School of Pure and Applied Sciences, University of Tsukuba,\\ Tsukuba, Ibaraki 305-8571, Japan} \end{center} 
\bigskip
\bigskip
\bigskip
\begin{abstract}  
We study the identity-based solution of Witten's cubic bosonic open string field theory constructed by Takahashi and Tanimoto, which is claimed to describe the tachyon vacuum. We argue that the observables of the solution coincide with those of the tachyon vacuum using the method proposed by Kishimoto and Takahashi. We also discuss how to treat the kinetic term of the  string field theory expanded around it.  
\end{abstract}

\setcounter{footnote}{0} \renewcommand{\thefootnote}{\arabic{footnote}}

\end{titlepage}

\section{Introduction}

Since the discovery of the tachyon vacuum solution by Schnabl \cite{Schnabl:2005gv},
various kinds of analytic solutions of the equation of motion of the
cubic bosonic open string field theory \cite{Witten:1985cc} have
been constructed (for reviews, see \cite{Fuchs:2008cc,Schnabl:2010tb,Okawa2012}).
It is now possible to construct a solution corresponding to any known
open string background \cite{Erler2014}. 

\begin{figure}[H]
\begin{centering}
\includegraphics[scale=0.5]{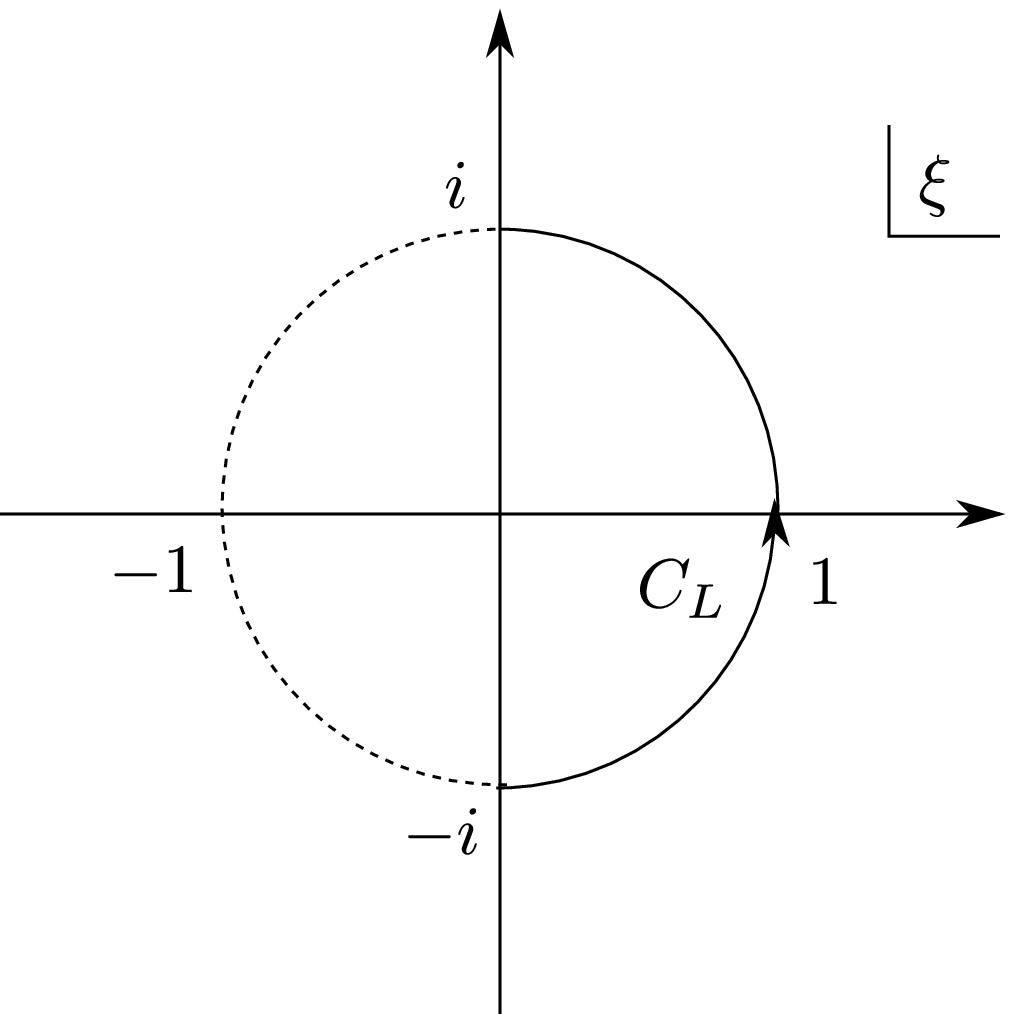}
\par\end{centering}

\caption{$C_{\mathrm{left}}$\label{fig:Cleft}}
\end{figure}

Most of the solutions found since \cite{Schnabl:2005gv} are so-called
regular solutions which consist mainly of wedge states with non vanishing
width with operator insertions. There exist some solutions which are
not of this kind. An example is the solution 
\begin{equation}
\Psi_{\mathrm{TT}}=\left[\int_{C_{\mathrm{left}}}\frac{d\xi}{2\pi i}\left(e^{h_{a}}-1\right)j_{\mathrm{B}}\left(\xi\right)-\int_{C_{\mathrm{left}}}\frac{d\xi}{2\pi i}\left(\partial h_{a}\right)^{2}e^{h_{a}}c\left(\xi\right)\right]\left|I\right\rangle \,,\label{eq:TTsolution}
\end{equation}
given by Takahashi and Tanimoto \cite{Takahashi:2002ez}, which is
called the scalar solution. Here $C_{\mathrm{left}}$ is a contour
in the upper half plane depicted in Fig. \ref{fig:Cleft}, $j_{\mathrm{B}}$
is the BRST current 
\begin{equation}
j_{\mathrm{B}}\left(\xi\right)=\left[cT+bc\partial c+\frac{3}{2}\partial^{2}c\right]\left(\xi\right)\,,
\label{eq:jbrst}
\end{equation}
$\left|I\right\rangle $ is the identity string field and $h_{a}\left(\xi\right)$
is a function taken to be 
\begin{equation}
h_{a}\left(\xi\right)=\ln\left(1+\frac{a}{2}\left(\xi+\frac{1}{\xi}\right)^{2}\right)\,,\label{eq:hxi}
\end{equation}
for $a\geq-\frac{1}{2}$. Takahashi and Tanimoto claim that while
the solution is a pure gauge solution for $a>-\frac{1}{2}$, it is
a tachyon vacuum solution for $a=-\frac{1}{2}$. 

The solution (\ref{eq:TTsolution}) is expressed as an identity state
with local operator insertions. The solutions of such a form are called
identity-based solutions. It is difficult to calculate observables
like energy or Ellwood invariant of identity-based solutions. These
quantities correspond to correlation functions of operators on a strip
with vanishing width in the worldsheet theory and naive regularizations
fail to yield definite values \cite{Kishimoto:2001de,Arroyo2010,Zeze2010}. 

On the other hand, the identity-based solutions have some advantages.
In general, the string field action expanded around a classical solution
$\Psi_{\mathrm{cl}}$ can be given as 
\begin{equation}
S^{\prime}\left[\Psi\right]=-\frac{1}{g^{2}}\int\left[\frac{1}{2}\Psi Q^{\prime}\Psi+\frac{1}{3}\Psi\Psi\Psi\right]\ ,\label{eq:Sprime}
\end{equation}
where
\[
Q^{\prime}A=QA+\Psi_{\mathrm{cl}}A-\left(-1\right)^{\left|\Psi_{\mathrm{cl}}\right|\left|A\right|}A\Psi_{\mathrm{cl}}\ .
\]
In the case of regular solutions, $\Psi_{\mathrm{cl}}$ involves wedge
states with finite width and it will be very difficult to study the
string field theory action (\ref{eq:Sprime}) with the kinetic operator
$Q^{\prime}$. However, if $\Psi_{\mathrm{cl}}$ is an identity-based
solution, the $Q^{\prime}$ can be expressed by local operators on
the worldsheet. For example, if $\Psi_{\mathrm{cl}}$ is the Takahashi-Tanimoto
solution (\ref{eq:TTsolution}), the $Q^{\prime}$ becomes
\begin{equation}
\oint\frac{d\xi}{2\pi i}\left[e^{h_{a}}j_{\mathrm{B}}\left(\xi\right)-\left(\partial h_{a}\right)^{2}e^{h_{a}}c\left(\xi\right)\right]\ .\label{eq:Qprime-1}
\end{equation}
With $Q^{\prime}$ being an operator like this, we expect it is relatively
easy to deal with the string field theory action (\ref{eq:Sprime}). 

Although the observables are not available, there are many evidences
indicating that the Takahashi-Tanimoto solution (\ref{eq:TTsolution})
with $a=-\frac{1}{2}$ is a tachyon vacuum solution:
\begin{itemize}
\item There are no physical open string excitations around the background
corresponding to $a=-\frac{1}{2}$. This fact has been shown by studying
the BRST cohomology \cite{Kishimoto:2002xi} or by constructing the
homotopy operator \cite{Inatomi2011}.
\item The open string amplitudes around the background can be shown to vanish
\cite{Takahashi:2003xe}. 
\item Solving the equation of motion in the background corresponding to
$a=-\frac{1}{2}$ numerically, an unstable solution which is supposed
to correspond to the perturbative vacuum can be found \cite{Takahashi2003b,Kishimoto:2009nd,Kishimoto2011}. 
\end{itemize}
All these evidences imply that the solution corresponds to the tachyon
vacuum. It should be interesting to explore the string field theory
around such a background and see whether or not the closed string
amplitudes can be reproduced from it. Since the solution is an identity-based
solution, the string field theory expanded around the solution will
have a tractable kinetic term. 

In this paper, we would like to study the Takahashi-Tanimoto solution
(\ref{eq:TTsolution}) with $a=-\frac{1}{2}$ and the string field
theory expanded around it. What we will do first is to evaluate the
observables of the solution in a rather indirect manner. In a recent
paper \cite{Inatomi2013a}, the authors consider the Erler-Schnabl
solutions in the string field theory expanded around the identity-based
marginal solutions found in \cite{Takahashi2001,Takahashi:2002ez}.
Since the Erler-Schnabl solutions will correspond to the tachyon vacuum,
by calculating the observables of these solutions, they are able to
evaluate the observables of the identity-based marginal solutions.
We here apply this method to the scalar solution (\ref{eq:TTsolution})
with $a=-\frac{1}{2}$ and see what we can say about the observables
of it. By doing so, we will get further evidences for the claim that
the solution is a tachyon vacuum solution. In the latter half of the
paper, we will discuss the string field theory expanded around 
the solution. We will show how we should treat the kinetic operator
(\ref{eq:Qprime-1}) in order for the solution to correspond to the
tachyon vacuum. 

The organization of this paper is as follows. In section \ref{sec:The-Erler-Schnabl-solution},
we evaluate the observables of the Takahashi-Tanimoto solution by
calculating those of the Erler-Schnabl solution in the string field
theory expanded around it. In section \ref{sec:String-field-theory},
we consider the string field theory around the Takahashi-Tanimoto
solution and discuss how we should treat the kinetic operator. Section
\ref{sec:Conclusions-and-discussions} is devoted to conclusions and
discussions. In appendix \ref{sec:Maccaferri's-method}, we discuss
the method proposed recently by Maccaferri \cite{Maccaferri2014}
to construct regular solutions gauge equivalent to identity-based
solutions. We explain what we can get by applying the method to the
solution (\ref{eq:TTsolution}). In appendix \ref{sec:Properties-of},
we derive some identities concerning the operators $U,U^{-1}$ which
play important roles in the main text.

\subsection*{Note added}

In the workshop ``String field theory and related aspects VI, SFT2014''
(July 28 -August 1, 2014, SISSA Italy), where this work is presented
\cite{Ishibashi-SFT2014}, we have learned that Kishimoto, Masuda
and Takahashi work on the same problem from a different point of view
\cite{Takahashi-SFT2014}\cite{Kishimoto-toappear}. Their results
have some overlap with those in section \ref{sec:The-Erler-Schnabl-solution}. 

While this paper was being typed, a paper \cite{Zeze2014a} appeared
on the arXiv, which also treat the same problem. There is some overlap
with the contents of appendix \ref{sec:Maccaferri's-method} but the
identity-based solution they deal with is different from ours.

\section{The Erler-Schnabl solution in the string field theory expanded around
the Takahashi-Tanimoto solution\label{sec:The-Erler-Schnabl-solution}}

\subsection{The Erler-Schnabl solution}

The Erler-Schnabl solution \cite{Erler:2009uj} 
\begin{equation}
\Psi_{\mathrm{ES}}=\frac{1}{1+K}\left(c+Q\left(Bc\right)\right)\,,\label{eq:ES}
\end{equation}
satisfies the equation of motion of the cubic string field theory.
Here $K,B,c$ are the string fields defined by
\begin{eqnarray*}
B & = & \int_{\frac{1}{2}-i\infty}^{\frac{1}{2}+i\infty}\frac{dz}{2\pi i}b\left(z\right)\left|I\right\rangle \,,\\
c & = & \left.c\left(z\right)\right|_{z=\frac{1}{2}}\left|I\right\rangle \,,\\
K & = & QB\\
 & = & \int_{\frac{1}{2}-i\infty}^{\frac{1}{2}+i\infty}\frac{dz}{2\pi i}T\left(z\right)\left|I\right\rangle \,,
\end{eqnarray*}
and the product of them is the star product. $z$ is the sliver frame
coordinate which is expressed by the upper half plane coordinate $\xi$
in (\ref{eq:TTsolution}) as
\[
z=\frac{2}{\pi}\arctan\xi\,.
\]
$K,B,c$ and $Q$ satisfy the so-called $KBc$ algebra \cite{Okawa:2006vm,Erler:2006hw}
and one can show that $\Psi_{\mathrm{ES}}$ is a solution by using
the algebra. The Erler-Schnabl solution $\Psi_{\mathrm{ES}}$ describes
the tachyon vacuum. This fact can be shown by calculating the observables
or by showing that 
\[
A=B\frac{1}{1+K}\,,
\]
gives the homotopy operator for the background $\Psi_{\mathrm{ES}}$,
i.e. $QA+\Psi_{\mathrm{ES}}A+A\Psi_{\mathrm{ES}}=1$ \cite{Ellwood:2006ba}.
The existence of the homotopy operator implies that there exist no
physical open string states around the background $\Psi_{\mathrm{ES}}$.

As was pointed out in \cite{Kishimoto2013b}, it is straightforward
to construct the Erler-Schnabl solution in the string field theory
(\ref{eq:Sprime}) expanded around an identity-based solution. $Q^{\prime}$
is a nilpotent operator and acts on string fields as a derivation.
It is easy to see that 
\begin{equation}
\Psi_{\mathrm{ES}}^{\prime}=\frac{1}{1+K^{\prime}}\left(c+Q^{\prime}\left(Bc\right)\right)\,.\label{eq:ESprime}
\end{equation}
with 
\[
K^{\prime}=Q^{\prime}B\,,
\]
satisfies the equation of motion derived from the string field action
(\ref{eq:Sprime}), because the $K^{\prime},B,c$ and $Q^{\prime}$
satisfy the same algebra as the $KBc$ and $Q$ do. Moreover, the
homotopy operator for the solution $\Psi_{\mathrm{ES}}^{\prime}$
can be constructed as
\[
A^{\prime}=B\frac{1}{1+K^{\prime}}\,.
\]
Therefore one can argue that the solution $\Psi_{\mathrm{ES}}^{\prime}$
describes the tachyon vacuum, provided $\frac{1}{1+K^{\prime}}$ is
a regular quantity. 

Let us consider the Erler-Schnabl solution $\Psi_{\mathrm{ES}}^{\prime}$
in the string field theory expanded around the Takahashi-Tanimoto
solution given in (\ref{eq:TTsolution}) with $a=-\frac{1}{2}$. In
this case, $Q^{\prime}$ is expressed by a contour integral
\begin{equation}
\oint\frac{dz}{2\pi i}\left[-\frac{\sin^{2}\pi z}{\cos^{2}\pi z}j_{\mathrm{B}}\left(z\right)+\frac{4\pi^{2}}{\cos^{4}\pi z}c\left(z\right)\right]\,,\label{eq:Qprime-2}
\end{equation}
in the sliver frame and $K^{\prime}$ becomes
\begin{eqnarray}
K^{\prime} & = & K+J\,,\nonumber \\
J
 & \equiv & 
\int_{\frac{1}{2}-i\infty}^{\frac{1}{2}+i\infty}\frac{dz}{2\pi i}\left[-\frac{1}{\cos^{2}\pi z}T^{\prime}\left(z\right)+\frac{4\pi^{2}}{\cos^{4}\pi z}\right]\left|I\right\rangle \,,\nonumber \\
T^{\prime}\left(z\right) & \equiv & T^{\mathrm{matter}}\left(z\right)-b\partial c\left(z\right)\,.\label{eq:Kprime}
\end{eqnarray}
$\frac{1}{1+K^{\prime}}$ can be expressed as 
\[
\frac{1}{1+K^{\prime}}=\int_{0}^{\infty}dLe^{-L\left(1+K^{\prime}\right)}\,,
\]
in the usual way and we need to define $e^{-LK^{\prime}}$ to make
sense of such quantities. In this section, we expand $e^{-LK^{\prime}}$
as 
\begin{eqnarray}
e^{-LK^{\prime}} & = & e^{-L\left(K+J\right)}\nonumber \\
 & = & \sum_{n=0}^{\infty}\left(-1\right)^{n}\lim_{\delta\to+0}\int_{\delta}^{\infty}dL_{1}\cdots\int_{\delta}^{\infty}dL_{n+1}\delta\left(\sum_{i=1}^{n+1}L_{i}-L\right)e^{-L_{1}K}Je^{-L_{2}K}J\cdots Je^{-L_{n+1}K}\,.\label{eq:perturbative}
\end{eqnarray}
and consider the right hand side as the definition of $e^{-LK^{\prime}}$.
From the point of view of the worldsheet theory, we define $e^{-LK^{\prime}}$
perturbatively treating $J$ as perturbation. 
The perturbation corresponds to adding
\begin{equation}
\int \frac{d^2z}{2\pi} \left[-\frac{1}{\cos^{2}\pi z}T^{\prime}\left(z\right)+\frac{4\pi^{2}}{\cos^{4}\pi z}\right]
\end{equation}
to the worldsheet action. 
Since it is a chiral quantity integrated over the bulk worldsheet, we do not encounter any ultraviolet divergences \cite{Inatomi2013a} 
and the expression is well-defined%
\footnote{Notice that the normalization of $J$ is fixed by the equation of
motion and there is no reason to expect that the higher order terms
in the expansion (\ref{eq:perturbative}) are small in any sense.
We will treat the operator $K^{\prime}$ without using such an expansion
in section \ref{sec:String-field-theory}.%
}. However, there is still a room for finite renormalizations. A prescription
for such renormalization is fixed by introducing a cut-off $\delta$. 

Now let us consider the observables of the Erler-Schnabl solution
$\Psi_{\mathrm{ES}}^{\prime}$. The observables we consider are the
action and the Ellwood invariant \cite{Hashimoto:2001sm,Gaiotto:2001ji,Ellwood:2008jh}.
The action of $\Psi_{\mathrm{ES}}^{\prime}$ in the string field theory
(\ref{eq:Sprime}) is equal to the difference of the energy between
the background corresponding to $\Psi_{\mathrm{TT}}$ and that corresponding
to $\Psi_{\mathrm{ES}}^{\prime}$. The Ellwood invariant of $\Psi_{\mathrm{ES}}^{\prime}$
becomes the difference of the $1$-point function of a closed string
vertex operator $V$ between these backgrounds. Thus they can be expressed
as
\begin{eqnarray}
S\left[\Psi_{\mathrm{ES}}^{\prime}\right] 
& = & 
E_{\mathrm{TT}}-E_{\Psi_{\mathrm{ES}}^{\prime}}\,,
\\
\mathrm{Tr}_{V}\Psi_{\mathrm{ES}}^{\prime} 
& = & 
\left\langle Vc\right\rangle _{\Psi_{\mathrm{ES}}^{\prime}}
-\left\langle Vc\right\rangle _{\mathrm{TT}}\,,
\end{eqnarray}
Here the Ellwood invariant $\mathrm{Tr}_{V}\Phi$ is given as
\begin{equation}
\mathrm{Tr}_{V}\Phi=\left\langle I\right|V\left(i,-i\right)\left|\Phi\right\rangle \,,
\label{Ellwood}
\end{equation}
 where $V\left(i,-i\right)=c\bar{c}V^{\mathrm{m}}\left(i,-i\right)$
is a closed string vertex operator.
$E_{\mathrm{TT}},E_{\Psi_{\mathrm{ES}}^{\prime}},
\left\langle Vc\right\rangle _{\Psi_{\mathrm{ES}}^{\prime}},
\left\langle Vc\right\rangle _{\mathrm{TT}}$ denote the
energy and the one-point function of each background respectively.
In the following, we will show $S\left[\Psi_{\mathrm{ES}}^{\prime}\right]=\mathrm{Tr}_{V}\Psi_{\mathrm{ES}}^{\prime}=0$,
which implies 
\begin{eqnarray}
E_{\Psi_{\mathrm{ES}}^{\prime}} 
& = &
 E_{\mathrm{TT}}\,,
\\
\left\langle Vc\right\rangle _{\Psi_{\mathrm{ES}}^{\prime}} 
& = & 
\left\langle Vc\right\rangle _{\mathrm{TT}}\,.
\end{eqnarray}
Since we assume that $\Psi_{\mathrm{ES}}^{\prime}$ corresponds to
the tachyon vacuum, we can see that the observables $E_{\mathrm{TT}},\left\langle Vc\right\rangle _{\mathrm{TT}}$
of the identity-based solution $\Psi_{\mathrm{TT}}$ coincide with
those of the tachyon vacuum. Therefore showing $S\left[\Psi_{\mathrm{ES}}^{\prime}\right]=\mathrm{Tr}_{V}\Psi_{\mathrm{ES}}^{\prime}=0$
gives evidences for the claim that the Takahashi-Tanimoto solution
$\Psi_{\mathrm{TT}}$ describes the tachyon vacuum.

In this section, we use this indirect way proposed in \cite{Inatomi2013a}
to calculate the observables $E_{\mathrm{TT}},\left\langle Vc\right\rangle _{\mathrm{TT}}$
of the identity-based solution $\Psi_{\mathrm{TT}}$. Recently there
are somewhat more direct ways to calculate these quantities \cite{Kishimoto2013b,Kishimoto2014}%
\footnote{Kishimoto, Masuda and Takahashi \cite{Kishimoto-toappear} generalize
the method of \cite{Kishimoto2013b,Kishimoto2014} to the case of
the scalar solutions. %
}\cite{Maccaferri2014}. Especially Maccaferri \cite{Maccaferri2014}
uses the so-called Zeze map \cite{Kishimoto2007} to construct regular
solutions gauge equivalent to identity-based ones and calculate the
observables of the regular ones. Moreover, the calculations eventually
reduce to those of the $S\left[\Psi_{\mathrm{ES}}^{\prime}\right],\mathrm{Tr}_{V}\Psi_{\mathrm{ES}}^{\prime}$.
In appendix \ref{sec:Maccaferri's-method}, we explain how we can
apply Maccaferri's method to the Takahashi-Tanimoto solution (\ref{eq:TTsolution})
with $a=-\frac{1}{2}$.

\subsection{Observables of $\Psi_{\mathrm{ES}}^{\prime}$}

Now let us calculate the observables $S\left[\Psi_{\mathrm{ES}}^{\prime}\right],\,\mathrm{Tr}_{V}\Psi_{\mathrm{ES}}^{\prime}$
and show that both of them vanish%
\footnote{Kishimoto, Masuda and Takahashi \cite{Kishimoto-toappear} obtain
the same results using a different method, considering more general
solutions made from $K^{\prime}Bc$.%
}. From the expression (\ref{eq:ESprime}), we obtain
\begin{eqnarray}
S\left[\Psi_{\mathrm{ES}}^{\prime}\right] & = & -\frac{1}{6g^{2}}\mathrm{Tr}\left[\frac{1}{1+K^{\prime}}c\frac{1}{1+K^{\prime}}Q^{\prime}c\right]\,,\nonumber \\
\mathrm{Tr}_{V}\Psi_{\mathrm{ES}}^{\prime} & = & \mathrm{Tr}_{V}\left[\frac{1}{1+K^{\prime}}c\right]\,.\label{eq:observables}
\end{eqnarray}
Therefore what we will prove are
\begin{eqnarray}
\mathrm{Tr}_{V}\left[\frac{1}{1+K^{\prime}}c\right] & = & 0\,,\label{eq:gioTT}\\
\mathrm{Tr}\left[\frac{1}{1+K^{\prime}}c\frac{1}{1+K^{\prime}}Q^{\prime}c\right] & = & 0\,.\label{eq:energyTT}
\end{eqnarray}

These can be proved by using the following identities:
\begin{eqnarray}
Q^{\prime}\left(\frac{1}{\pi^{2}}b\right) & = & 1\,,\label{eq:homotopy}\\
Q^{\prime}c & = & 0\,.\label{eq:Qprimec}
\end{eqnarray}
Here
\[
\frac{1}{\pi^{2}}b\equiv\left.\frac{1}{\pi^{2}}b\left(z\right)\right|_{z=\frac{1}{2}}\left|I\right\rangle =\left.b\left(\xi\right)\right|_{\xi=1}\left|I\right\rangle \,,
\]
and (\ref{eq:homotopy}) suggests that the $\frac{1}{\pi^{2}}b$ works
as a homotopy operator of the BRST charge $Q^{\prime}$. 

\begin{figure}
\begin{centering}
\includegraphics[scale=0.5]{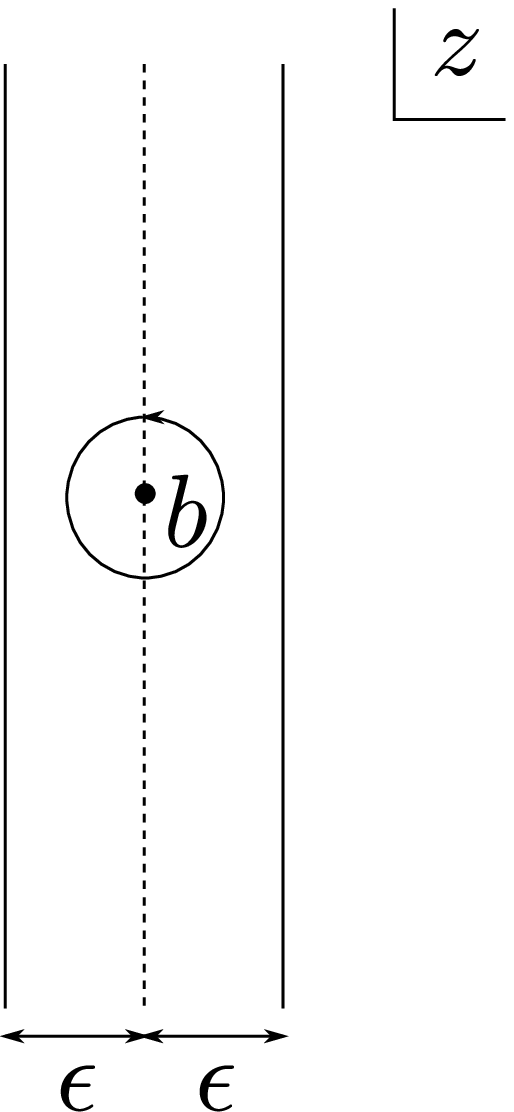}
\par\end{centering}

\caption{$e^{-\epsilon K}Q^{\prime}be^{-\epsilon K}$\label{fig:Eq.homotopy}}
\end{figure}

To be precise, one can show (\ref{eq:homotopy})(\ref{eq:Qprimec})
in the situation where we have some worldsheet around $\frac{1}{\pi^{2}}b,c$
without any local operator insertions. Namely we should consider 
\begin{eqnarray}
e^{-\epsilon K}Q^{\prime}\left(\frac{1}{\pi^{2}}b\right)e^{-\epsilon K} & = & e^{-2\epsilon K}\,,\label{eq:homotopy2}\\
e^{-\epsilon K}Q^{\prime}ce^{-\epsilon K} & = & 0\,,\label{eq:Qprimec2}
\end{eqnarray}
in which we attach $e^{-\epsilon K}$'s to generate worldsheet as
is depicted in Fig. \ref{fig:Eq.homotopy}. %
\footnote{In \cite{Maccaferri2014}, such a prescription is used for the equation
of motion. One can show that the $\Psi_{\mathrm{TT}}$ in (\ref{eq:TTsolution})
satisfies the equation of motion in the same way.%
}. With the worldsheet, one can express the action of $Q^{\prime}$
by the contour integral (\ref{eq:Qprime-2}) and get
\begin{eqnarray}
 &  & 
e^{-\epsilon K}Q^{\prime}\left(\frac{1}{\pi^{2}}b\right)e^{-\epsilon K}
\nonumber \\
&  & 
\quad=e^{-\epsilon K}\left(\oint_{0}\frac{dz}{2\pi i}\left[-\frac{\sin^{2}\pi z}{\cos^{2}\pi z}j_{\mathrm{B}}\left(z\right)+\frac{4\pi^{2}}{\cos^{4}\pi z}c\left(z\right)\right]\frac{1}{\pi^{2}}b\left(0\right)\right)e^{-\epsilon K}
\nonumber \\
&  & 
\quad=e^{-\epsilon K}\left(\oint_{0}\frac{dz}{2\pi i}\left[-\frac{\sin^{2}\pi z}{\cos^{2}\pi z}
\left(\frac{3}{2}\partial^2c\left(z\right)\right)
+\frac{4\pi^{2}}{\cos^{4}\pi z}c\left(z\right)\right]\frac{1}{\pi^{2}}b\left(0\right)\right)e^{-\epsilon K}
\nonumber \\
&  & 
\quad=e^{-2\epsilon K}\,.
\label{eq:homotopy3}
\end{eqnarray}
(\ref{eq:Qprimec2}) can be derived in the same way.

(\ref{eq:energyTT}) is an immediate consequence of  (\ref{eq:Qprimec2}).
(\ref{eq:gioTT}) can be derived from (\ref{eq:homotopy2})(\ref{eq:Qprimec2})
as follows. Inserting (\ref{eq:homotopy}) into $\mathrm{Tr}_{V}\left[\frac{1}{1+K^{\prime}}c\right]$,
we get
\[
\mathrm{Tr}_{V}\left[\frac{1}{1+K^{\prime}}c\right]=\mathrm{Tr}_{V}\left[\frac{1}{\sqrt{1+K^{\prime}}}Q^{\prime}\left(\frac{1}{\pi^{2}}b\right)\frac{1}{\sqrt{1+K^{\prime}}}c\right]\,.
\]
Here we use the definition 
\[
\frac{1}{\sqrt{1+K^{\prime}}}=\frac{1}{\Gamma\left(\frac{1}{2}\right)}\int_{0}^{\infty}dLL^{-\frac{1}{2}}e^{-L}e^{-LK^{\prime}}\,,
\]
where $e^{-LK^{\prime}}$ expressed as (\ref{eq:perturbative}). With
the cutoff $\delta$, (\ref{eq:homotopy})(\ref{eq:Qprimec}) can
be safely used because there are some worldsheets with no operator
insertions around $b,c$. Thus we obtain
\begin{eqnarray*}
 &  & \mathrm{Tr}_{V}\left[\frac{1}{1+K^{\prime}}c\right]\\
 &  & \quad=\mathrm{Tr}_{V}\left[\frac{1}{\sqrt{1+K^{\prime}}}\frac{1}{\pi^{2}}b\frac{1}{\sqrt{1+K^{\prime}}}Q^{\prime}c\right]\\
 &  & \hphantom{\quad}=0\,.
\end{eqnarray*}

Before closing this section, one comment is in order. Using $Q^{\prime}c=0$,
one can see that from (\ref{eq:ESprime}) 
\[
\Psi_{\mathrm{ES}}^{\prime}=c\,.
\]
Thus actually the $\Psi_{\mathrm{ES}}^{\prime}$ itself is an identity-based
solution%
\footnote{One may be able to calculate the observables for such a solution following
\cite{Erler2012} or \cite{Zeze2014a}.%
}, although we do not have any trouble in calculating the right hand
sides of (\ref{eq:observables}). One can avoid this by replacing
$c$ by
\[
c_{y}\equiv c\left(\frac{1}{2}+iy\right)\left|I\right\rangle \,\left(y\ne0,\ y\in\mathbb{R}\right)\,.
\]
$K^{\prime},B,c_{y}$ satisfy the $KBc$ algebra and one can construct
the Erler-Schnabl solution 
\[
\Psi_{\mathrm{ES},y}^{\prime}=\frac{1}{1+K^{\prime}}\left(c_{y}+Q^{\prime}\left(Bc_{y}\right)\right)\,,
\]
which is not identity-based, albeit it still includes an identity based piece. 
The observables to be calculated become
\begin{eqnarray}
S\left[\Psi_{\mathrm{ES},y}^{\prime}\right] & = & -\frac{1}{6g^{2}}\mathrm{Tr}\left[\frac{1}{1+K^{\prime}}c_{y}\frac{1}{1+K^{\prime}}Q^{\prime}c_{y}\right]\,,\nonumber \\
\mathrm{Tr}_{V}\Psi_{\mathrm{ES},y}^{\prime} & = & \mathrm{Tr}_{V}\left[\frac{1}{1+K^{\prime}}c_{y}\right]\,.\label{eq:ESprimey}
\end{eqnarray}
One can show that these quantities are actually independent of $y$.
Indeed, using the $KBc$ identity
\[
\left\{ B,c_{y}\right\} =1\,,
\]
the formulas given in \cite{Maccaferri2014} (eqs.(3.4),(3.10)-(3.19))
imply
\begin{eqnarray*}
 &  & \frac{1}{6}\mathrm{Tr}\left[\frac{1}{1+K^{\prime}}c_{y}\frac{1}{1+K^{\prime}}Q^{\prime}c_{y}\right]\\
 &  & \quad=\frac{1}{6}\mathrm{Tr}\left[\frac{1}{1+K}c_{y}\frac{1}{1+K}Qc_{y}\right]-\frac{1}{6}\mathrm{Tr}\left[\frac{1}{1+K}\Psi_{\mathrm{TT}}\frac{1}{1+K^{\prime}}\Psi_{\mathrm{TT}}\frac{1}{1+K}\Psi_{\mathrm{TT}}\frac{1}{1+K^{\prime}}\right]\,,\\
 &  & \mathrm{Tr}_{V}\left[\frac{1}{1+K^{\prime}}c_{y}\right]\\
 &  & \quad=\mathrm{Tr}_{V}\left[\frac{1}{1+K}c_{y}\right]-\mathrm{Tr}_{V}\left[\frac{1}{1+K}\Psi_{\mathrm{TT}}\frac{1}{1+K^{\prime}}\right]\,,
\end{eqnarray*}
and the right hand sides are independent of $y$. Therefore evaluating
them at $y=0$, we can see that the observables (\ref{eq:ESprimey})
all vanish.

\section{String field theory expanded around the Takahashi-Tanimoto solution\label{sec:String-field-theory}}

The derivation in the previous section uses the perturbative definition
(\ref{eq:perturbative}) of $e^{-LK^{\prime}}$. Since $\Psi_{\mathrm{TT}}$
is an identity-based solution, the kinetic term $Q^{\prime}$ is given
by an integral of local operators on the worldsheet and we should
be able to treat $K^{\prime}$ more directly. In this section, we
will examine if we can derive the results in section \ref{sec:The-Erler-Schnabl-solution}
by doing so. 

In the calculation of the observables in the previous section, the
following relations were essential:
\begin{eqnarray*}
e^{-\epsilon K}Q^{\prime}\left(\frac{1}{\pi^{2}}b\right)e^{-\epsilon K} & = & e^{-2\epsilon K}\,,\\
e^{-\epsilon K}Q^{\prime}ce^{-\epsilon K} & = & 0\,.
\end{eqnarray*}
These relations hold for the perturbative definition of $e^{-LK^{\prime}}$.
In the treatment here, it will be more appropriate to consider 
\begin{eqnarray}
e^{-\epsilon K^{\prime}}Q^{\prime}\left(\frac{1}{\pi^{2}}b\right)e^{-\epsilon K^{\prime}} & = & e^{-2\epsilon K^{\prime}}\,,\label{eq:Qb}\\
e^{-\epsilon K^{\prime}}Q^{\prime}ce^{-\epsilon K^{\prime}} & = & 0\,,\label{eq:Qc}
\end{eqnarray}
where the $e^{-\epsilon K^{\prime}}$'s are expected to provide worldsheet
with no operator insertions. 

Actually, as we will see, the definition of $e^{-LK^{\prime}}$ is
very subtle and we need some regularization to define quantities involving
it. There seem to be many ways to treat it, which should be related
to the choice of the prescription of renormalization in the perturbative
definition of $e^{-LK^{\prime}}$ (\ref{eq:perturbative}). Here we
use the identities (\ref{eq:Qb})(\ref{eq:Qc}) and their consequences
(\ref{eq:gioTT})(\ref{eq:energyTT}) as the guiding principle to
find the definition of $e^{-LK^{\prime}}$ so that the string field
action (\ref{eq:Sprime}) should describe the tachyon vacuum.

\subsection{Similarity transformation}

The $K^{\prime}$ given in (\ref{eq:Kprime}) involves $T^{\prime}\left(z\right)$
which is a twisted energy momentum tensor with central charge $c=24$.
Therefore we need to take care of the conformal anomaly on the worldsheet
to deal with the correlation functions on surfaces generated by $e^{-LK^{\prime}}$
and the calculations will become cumbersome. Here we would like to
use an alternative way of dealing with $K^{\prime}$ to do calculations. 

As was pointed out by Kishimoto and Takahashi \cite{Kishimoto:2002xi},
the kinetic operator $Q^{\prime}$ of the string field theory expanded
around the solution (\ref{eq:TTsolution}) with $a=-\frac{1}{2}$
can be expressed as

\begin{equation}
Q^{\prime}=e^{-q}\left(-\frac{1}{4}Q_{2}+c_{2}\right)e^{q}\,,\label{eq:Qprime}
\end{equation}
where
\begin{eqnarray}
q
& = & 
-\oint\frac{d\xi}{2\pi i}
\left(-bc\right)\left(\xi\right)
\ln\left(1-\frac{1}{\xi^{2}}\right)^{2}\,,
\\
Q_{k} 
& = & 
\oint\frac{d\xi}{2\pi i}\xi^{k}j_{\mathrm{B}}\left(\xi\right)\,,
\\
c_{k} 
& = & 
\oint\frac{d\xi}{2\pi i}\xi^{k-2}c\left(\xi\right)\,.
\end{eqnarray}
Using the mode expansion of the ghost number current 
\[
-bc\left(\xi\right)=\sum_{n}j_{n}\xi^{-n-1}\,,
\]
the $q$ is expressed as 
\[
q=2\sum_{n=1}^{\infty}\frac{1}{n}j_{-2n}\,.
\]

\subsection*{$bc$-shift operation}

Eq. (\ref{eq:Qprime}) can be rewritten by using the $bc$-shift operation
\cite{Kishimoto:2002xi} defined for $k\in\mathbb{Z}$ as 
\begin{eqnarray*}
c_{n} & \to & c_{n}^{\left(k\right)}=c_{n+k}\,,\\
b_{n} & \to & b_{n}^{\left(k\right)}=b_{n-k}\,,\\
\left|0\right\rangle  & \to & \left|0\right\rangle ^{\left(k\right)}=\begin{cases}
b_{-k-1}b_{-k}\cdots b_{-2}\left|0\right\rangle  & k>0\\
c_{k+2}c_{k+3}\cdots c_{1}\left|0\right\rangle  & k<0
\end{cases}\,,\\
\left\langle 0\right| & \to & ^{\left(k\right)}\negmedspace\left\langle 0\right|=\begin{cases}
\left\langle 0\right|c_{-1}c_{0}\cdots c_{k-2} & k>0\\
\left\langle 0\right|b_{2}b_{3}\cdots b_{-k+1} & k<0
\end{cases}\,,
\end{eqnarray*}
and $\phi\to\phi^{\left(k\right)}=\phi$ if $\phi$ involves only
matter fields. A state 
\[
\left|a\right\rangle =\phi_{-n_{1}}\cdots b_{-m_{1}}\cdots c_{-l_{1}}\cdots\left|0\right\rangle \,,
\]
in the Fock space is mapped to 
\[
\left|a\right\rangle ^{\left(k\right)}=\phi_{-n_{1}}^{\left(k\right)}\cdots b_{-m_{1}}^{\left(k\right)}\cdots c_{-l_{1}}^{\left(k\right)}\cdots\left|0\right\rangle ^{\left(k\right)}\,,
\]
under this operation. $c_{n}^{\left(k\right)},b_{n}^{\left(k\right)},\left|0\right\rangle ^{\left(k\right)},{}^{\left(k\right)}\negmedspace\left\langle 0\right|$
satisfy
\begin{eqnarray}
\left\{ c_{n}^{\left(k\right)},b_{n}^{\left(h\right)}\right\}  & = & \delta_{n+m,0}\,,\nonumber \\
\left\{ c_{n}^{\left(k\right)},c_{m}^{\left(k\right)}\right\} =\left\{ b_{n}^{\left(k\right)},b_{m}^{\left(k\right)}\right\}  & = & 0\,,\nonumber \\
b_{n}^{\left(k\right)}\left|0\right\rangle ^{\left(k\right)} & = & 0\,\left(n\geq-1\right)\,,\nonumber \\
c_{n}^{\left(k\right)}\left|0\right\rangle ^{\left(k\right)} & = & 0\,\left(n\geq2\right)\,,\nonumber \\
^{\left(k\right)}\negmedspace\left\langle 0\right|b_{n}^{\left(k\right)} & = & 0\,\left(n\leq1\right)\,,\nonumber \\
^{\left(k\right)}\negmedspace\left\langle 0\right|c_{n}^{\left(k\right)} & = & 0\,\left(n\leq-2\right)\,,\label{eq:bcrelation1}
\end{eqnarray}
and 
\begin{eqnarray}
 &  & ^{\left(k\right)}\negmedspace\left\langle 0\right|c_{-1}^{\left(k\right)}c_{0}^{\left(k\right)}c_{1}^{\left(k\right)}\left|0\right\rangle ^{\left(k\right)}\nonumber \\
 &  & \quad=\left\langle 0\right|c_{-1}c_{0}c_{1}\left|0\right\rangle \nonumber \\
 &  & \quad=1\,.\label{eq:bcrelation2}
\end{eqnarray}
Since we can evaluate all the correlation functions of the $bc$ system
using the relations (\ref{eq:bcrelation1})(\ref{eq:bcrelation2}),
we can see that for any states $\left\langle a\right|,\left|b\right\rangle $ in the Fock space,
\[
^{\left(k\right)}\negmedspace\left\langle a|b\right\rangle ^{\left(k\right)}=\left\langle a|b\right\rangle \,.
\]
Under the $bc$-shift operation, the BRST charge is transformed as
\[
Q\to Q^{\left(k\right)}=Q_{k}-k^{2}c_{k}\,.
\]
Therefore (\ref{eq:Qprime}) can be written as 
\begin{equation}
Q^{\prime}=-\frac{1}{4}e^{-q}Q^{\left(2\right)}e^{q}\,.\label{eq:Qprime2}
\end{equation}

It is convenient to introduce operators $U_{k}\ \left(k\in\mathbb{Z}\right)$
which are defined so that
\begin{eqnarray*}
U_{k}\left|a\right\rangle  & = & \left|a\right\rangle ^{\left(k\right)}\,,\\
\left\langle a\right|U_{k} & = & ^{\left(-k\right)}\negmedspace\left\langle a\right|\,.
\end{eqnarray*}
$U_{k}$ satisfies
\begin{eqnarray*}
U_{k}U_{-k}\left|a\right\rangle 
& = &
\left|a\right\rangle
 \,,\\
U_{k}\mathcal{O}U_{-k} & = & \mathcal{O}^{\left(k\right)}\,,
\end{eqnarray*}
for any state $\left|a\right\rangle$ in the Fock space and any operator $\mathcal{O}$. 
It turns out that $U_{k}$ can be expressed as
\begin{equation}
U_{k}
=
e^{-k\sigma_{0}}
\,,
\label{eq:Uk}
\end{equation}
where $\sigma_{0}$ is the operator which appears in the bosonization
formulas (\ref{eq:cbosonization}), (\ref{eq:bbosonization}).
Indeed, $e^{-k\sigma_{0}}$ satisfies
\begin{eqnarray*}
e^{-k\sigma_{0}}c\left(\xi\right)e^{k\sigma_{0}} & = & e^{-k\sigma_{0}}\exp\left[\sum_{n=1}^{\infty}\frac{1}{n}j_{-n}\xi^{n}\right]e^{\sigma_{0}}e^{j_{0}\ln\xi}\exp\left[-\sum_{n=1}^{\infty}\frac{1}{n}j_{n}\xi^{-n}\right]e^{k\sigma_{0}}=\xi^{k}c\left(\xi\right)\,,\\
e^{-k\sigma_{0}}b\left(\xi\right)e^{k\sigma_{0}} & = & e^{-k\sigma_{0}}\exp\left[-\sum_{n=1}^{\infty}\frac{1}{n}j_{-n}\xi^{n}\right]e^{-\sigma_{0}}e^{-j_{0}\ln\xi}\exp\left[\sum_{n=1}^{\infty}\frac{1}{n}j_{n}\xi^{-n}\right]e^{k\sigma_{0}}=\xi^{-k}b\left(\xi\right)\,,\\
e^{-k\sigma_{0}}\left|0\right\rangle  & = & \begin{cases}
b_{-k-1}b_{-k}\cdots b_{-2}\left|0\right\rangle =\left|0\right\rangle ^{\left(k\right)} & k>0\\
c_{k+2}c_{k+3}\cdots c_{1}\left|0\right\rangle =\left|0\right\rangle ^{\left(k\right)} & k<0
\end{cases}\,,\\
\left\langle 0\right|e^{-k\sigma_{0}} & = & \begin{cases}
\left\langle 0\right|b_{2}b_{3}\cdots b_{k+1}=^{\left(-k\right)}\left\langle 0\right| & k>0\\
\left\langle 0\right|c_{-1}c_{0}\cdots c_{-k-2}=^{\left(-k\right)}\left\langle 0\right| & k<0
\end{cases}\,.
\end{eqnarray*}
From (\ref{eq:Uk}) and 
\[
\left[j_{0},e^{-k\sigma_{0}}\right]=-ke^{-k\sigma_{0}}\,,
\]
we can see that $U_{k}$ carries ghost number $-k$.


(\ref{eq:Qprime2}) can be written
as 
\begin{equation}
Q^{\prime}=-\frac{1}{4}UQU^{-1}\,.\label{eq:Qprime3}
\end{equation}
where
\begin{eqnarray}
U & \equiv & e^{-q}U_{2}\,,\nonumber \\
U^{-1} & \equiv & U_{-2}e^{q}\,.\label{eq:U}
\end{eqnarray}
Notice that $U,U^{-1}$ are of ghost number $-2,2$. 
$U$ and $U^{-1}$ are inverse to each other, when these operators act on the states 
in the Fock space. 
However, when we are dealing with the states outside of the Fock space, 
such a statement may become subtle, as is discussed in appendix 
\ref{sec:Properties-of}.  
Another thing
to be noticed is that the BPZ conjugates of $U,U^{-1}$ do not coincide
with either $U$ or $U^{-1}$. 

Therefore, the $Q^{\prime}$ is related to the original kinetic operator
$Q$ by a similarity transformation (\ref{eq:Qprime3}), which implies
that the solution $\Psi_{\mathrm{TT}}$ is formally in the pure gauge
form. By the similarity transformation, $K^{\prime}$ is turned into
an operator made from $T=\left\{ Q,b\right\} $ and thus it is possible
to evaluate quantities involving $K^{\prime}$ without dealing with
the twisted energy momentum tensor $T^{\prime}$.

\subsection{$U,U^{-1}$ \label{subsec:UU-1}}

We need some identities satisfied by $U,U^{-1}$ to perform calculations
using the relation (\ref{eq:Qprime3}). From the definition (\ref{eq:U})
we obtain
\begin{eqnarray}
Uc\left(\xi\right)U^{-1} & = & \frac{\left(\xi^{2}-1\right)^{2}}{\xi^{2}}c\left(\xi\right)=-4e^{h_{-\frac{1}{2}}\left(\xi\right)}c\left(\xi\right)\,,\label{eq:UcU-1}\\
U^{-1}c\left(\xi\right)U & = & \frac{\xi^{2}}{\left(\xi^{2}-1\right)^{2}}c\left(\xi\right)=-\frac{1}{4}e^{-h_{-\frac{1}{2}}\left(\xi\right)}c\left(\xi\right)\,,\label{eq:U-1cU}\\
Ub\left(\xi\right)U^{-1} & = & \frac{\xi^{2}}{\left(\xi^{2}-1\right)^{2}}b\left(\xi\right)=-\frac{1}{4}e^{-h_{-\frac{1}{2}}\left(\xi\right)}b\left(\xi\right)\,,\label{eq:UbU-1}\\
U^{-1}b\left(\xi\right)U & = & \frac{\left(\xi^{2}-1\right)^{2}}{\xi^{2}}b\left(\xi\right)=-4e^{h_{-\frac{1}{2}}\left(\xi\right)}c\left(\xi\right)\,.\label{eq:U-1bU}
\end{eqnarray}
It is also possible to derive how $U,U^{-1}$ act on the states $\left|0\right\rangle ,\left|I\right\rangle ,\left\langle 0\right|,\left\langle I\right|$:
\begin{eqnarray}
U\left|0\right\rangle  & = & \frac{1}{16}\partial bb\left(1\right)\partial bb\left(-1\right)c_{0}c_{1}\left|0\right\rangle \,,\label{eq:U0}\\
U^{-1}\left|0\right\rangle  & = & \frac{1}{16}\partial cc\left(1\right)\partial cc\left(-1\right)b_{-3}b_{-2}\left|0\right\rangle \,,\label{eq:U-10}\\
\left\langle 0\right|U & = & \left\langle 0\right|b_{2}b_{3}\,,\label{eq:0U}\\
\left\langle 0\right|U^{-1} & = & \left\langle 0\right|c_{-1}c_{0}\,.\label{eq:0U-1}\\
U\left|I\right\rangle  & = & \frac{1}{32}\partial bb\left(1\right)\left|I\right\rangle \,,\label{eq:UI}\\
U^{-1}\left|I\right\rangle  & = & 2\partial cc\left(1\right)\left|I\right\rangle \,.\label{eq:U-1I}
\end{eqnarray}
Moreover, one can show that 
 $\left\langle I\right|U$ and $\left\langle I\right|U^{-1}$ can be set to zero 
in the situations where no ghost operators are inserted at $\xi =\pm 1$. 
These properties are proved in appendix \ref{sec:Properties-of}. 

Here let us comment on one thing concerning the operators $U,U^{-1}$, 
which will be relevant to the subsequent discussions. 
The pure gauge form (\ref{eq:Qprime3}) apparently contradicts the
existence of the homotopy operator (\ref{eq:homotopy}), as was pointed
out in \cite{Inatomi2011,Inatomi2011a}. Indeed, one can see from
(\ref{eq:Qprime3}) that the representatives of the BRST cohomology
of $Q^{\prime}$ are given by the states of the form \cite{Kishimoto:2002xi}
\begin{eqnarray}
UcV^{\mathrm{m}}\left(0\right)\left|0\right\rangle  & : & \mathrm{gh\#}=-1\,,\label{eq:ghn-1}\\
U\partial ccV^{\mathrm{m}}\left(0\right)\left|0\right\rangle  & : & \mathrm{gh\#}=0\,.\label{eq:ghn0}
\end{eqnarray}
where $V^{\mathrm{m}}$ is a primary field made from the matter fields
with weight $1$. Therefore one can conclude that there exist no physical
open string excitations because they correspond to the states with
ghost number $1$. On the other hand, the existence of the homotopy
operator $b\left(1\right)$ implies that the states (\ref{eq:ghn-1})(\ref{eq:ghn0})
should be written in a BRST exact form
\begin{eqnarray*}
UcV^{\mathrm{m}}\left(0\right)\left|0\right\rangle  & = & Q^{\prime}b\left(1\right)UcV^\mathrm{m}\left(0\right)\left|0\right\rangle \,,\\
U\partial ccV^{\mathrm{m}}\left(0\right)\left|0\right\rangle  & = & Q^{\prime}b\left(1\right)U\partial ccV^\mathrm{m}\left(0\right)\left|0\right\rangle \,.
\end{eqnarray*}
Actually these do not hold. Indeed, using eqs.(\ref{eq:U-1cU})(\ref{eq:U0}),
we obtain
\begin{eqnarray}
UcV^{\mathrm{m}}\left(0\right)\left|0\right\rangle  & = & \frac{1}{16}\partial bb\left(1\right)\partial bb\left(-1\right)c_{-1}c_{0}c_{1}V^{\mathrm{m}}\left(0\right)\left|0\right\rangle \,,\label{eq:UcV0}\\
U\partial ccV^{\mathrm{m}}\left(0\right)\left|0\right\rangle  & = & \frac{1}{16}\partial bb\left(1\right)\partial bb\left(-1\right)c_{-2}c_{-1}c_{0}c_{1}V^{\mathrm{m}}\left(0\right)\left|0\right\rangle \,,\label{eq:UdccV0}
\end{eqnarray}
and $b\left(1\right)UcV\left(0\right)\left|0\right\rangle =b\left(1\right)U\partial ccV\left(0\right)\left|0\right\rangle =0$.
The reason for this apparent contradiction is that the relation (\ref{eq:homotopy})
holds only when there is some worldsheet around $b\left(1\right)$
without any local operator insertions, as we mentioned below eq.(\ref{eq:Qprimec2}).
Therefore, for the states (\ref{eq:UcV0})(\ref{eq:UdccV0}) which
involve $\partial bb\left(1\right)$, $b\left(1\right)$ does not
work as a homotopy operator of $Q^{\prime}$. 

\subsection{Calculations of the observables}

Now we would like to discuss how we can evaluate the observables (\ref{eq:observables})
using the expression (\ref{eq:Qprime3}). In order to facilitate the
calculation using eq.(\ref{eq:Qprime3}), we rewrite everything in
terms of the first-quantized operators, rather than string fields.
Here let us introduce $\mathcal{B}^{+},\mathcal{L}^{\prime+}$ such
that \cite{Schnabl:2002gg,Schnabl:2005gv} 
\begin{eqnarray}
\mathcal{B}^{+} & = & \oint\frac{d\xi}{2\pi i}\left(1+\xi^{2}\right)\left(\tan^{-1}\xi+\tan^{-1}\left(\frac{1}{\xi}\right)\right)b\left(\xi\right)\nonumber \\
 & = & \frac{\pi}{2}\oint\frac{d\xi}{2\pi i}\left(1+\xi^{2}\right)\epsilon\left(\mathrm{Re}\xi\right)b\left(\xi\right)\,,\nonumber \\
\mathcal{L}^{\prime+} & \equiv & \left\{ Q^{\prime},\mathcal{B}^{+}\right\} \,.\label{eq:Lprime+}
\end{eqnarray}
$\mathcal{L}^{\prime+}$ is the translation operator with respect
to the sliver frame coordinate $z$ for the left and right half of
the string. Therefore, the action of $\mathcal{L}^{\prime+}$ on any state $|\phi\rangle$ 
can be expressed by the string field $K^{\prime}$ as
\begin{equation}
\mathcal{L}^{\prime+}|\phi \rangle
=
K^{\prime}\ast |\phi \rangle
+
|\phi \rangle\ast K^{\prime}\,.
\label{curlyLprime+}
\end{equation}
 $\mathcal{L}^{\prime+}$ can be used to express various quantities involving 
$K^\prime$ in our setup. 
For example, using (\ref{curlyLprime+}) and (\ref{Ellwood}), one can show that 
\begin{eqnarray}
& &
\left\langle I\right|e^{-\frac{L}{4}\mathcal{L}^{\prime+}}c\left(1\right)V\left(i,-i,\right)e^{-\frac{L}{4}\mathcal{L}^{\prime+}}\left|I\right\rangle
\nonumber
\\
& &
\quad
=\mathrm{Tr}_V
\left[
e^{-\frac{L}{4}K^\prime}\ast c\ast e^{-\frac{L}{4}K^\prime}\ast 
\left|I\right\rangle\ast e^{-\frac{L}{4}K^\prime}\ast e^{-\frac{L}{4}K^\prime}
\right]
\nonumber
\\
& &
\quad
=\mathrm{Tr}_{V}\left[e^{-LK^{\prime}}c\right] \,,
\end{eqnarray}
holds and the left hand side of eqs.(\ref{eq:gioTT}) is expressed as 
\begin{equation}
\mathrm{Tr}_{V}\left[\frac{1}{1+K^{\prime}}c\right] 
=
\int_{0}^{\infty}dLe^{-L}
\left\langle I\right|e^{-\frac{L}{4}\mathcal{L}^{\prime+}}c\left(1\right)V\left(i,-i,\right)e^{-\frac{L}{4}\mathcal{L}^{\prime+}}\left|I\right\rangle \,.
\label{eq:gioprime}
\end{equation}
In a similar way, one gets
\begin{eqnarray}
& &
\left\langle I\right|e^{-\frac{L_{1}-L_{2}}{2}\mathcal{L}^{\prime+}}c\left(1\right)e^{-L_{2}\mathcal{L}^{\prime+}}Q^{\prime}c\left(1\right)\left|I\right\rangle 
\nonumber
\\
& &
\quad 
=\mathrm{Tr}\left[e^{-\frac{L_{1}-L_{2}}{2}K^{\prime}}\ast c\ast 
e^{-L_2K^{\prime}}\ast Q^{\prime}c \ast e^{-L_2K^{\prime}}\ast e^{-\frac{L_{1}-L_{2}}{2}K^{\prime}}
\right]
\nonumber
\\
& &
\quad
=\mathrm{Tr}\left[e^{-L_1K^{\prime}}ce^{-L_2K^{\prime}}Q^{\prime}c\right]
\,.
\end{eqnarray}
We expect that $e^{-\frac{L_{1}-L_{2}}{2}\mathcal{L}^{\prime+}}$ is well-defined when $L_1>L_2$ and this 
equation is valid only for $L_1>L_2$. When $L_2>L_1$, 
$\left\langle I\right|e^{-\frac{L_{2}-L_{1}}{2}\mathcal{L}^{\prime+}}c\left(-1\right)e^{-L_{1}\mathcal{L}^{\prime+}}Q^{\prime}c\left(1\right)\left|I\right\rangle$ can be used to express 
$\mathrm{Tr}\left[e^{-L_1K^{\prime}}ce^{-L_2K^{\prime}}Q^{\prime}c\right]$.
Therefore
the left hand side
of (\ref{eq:energyTT}) is expressed as
\begin{eqnarray}
\mathrm{Tr}\left[\frac{1}{1+K^{\prime}}c\frac{1}{1+K^{\prime}}Q^{\prime}c\right] 
&=&
\int_{0}^{\infty}dL_{2}e^{-L_{2}}\int_{L_{2}}^{\infty}dL_{1}e^{-L_{1}}
\nonumber 
\\
 &  & 
\qquad\times\left\langle I\right|e^{-\frac{L_{1}-L_{2}}{2}\mathcal{L}^{\prime+}}c\left(1\right)e^{-L_{2}\mathcal{L}^{\prime+}}Q^{\prime}c\left(1\right)\left|I\right\rangle 
\nonumber \\
 &  & 
+\int_{0}^{\infty}dL_{2}e^{-L_{2}}\int_{0}^{L_{2}}dL_{1}e^{-L_{1}}
\nonumber \\
 &  & 
\qquad\times\left\langle I\right|e^{-\frac{L_{2}-L_{1}}{2}\mathcal{L}^{\prime+}}c\left(-1\right)e^{-L_{1}\mathcal{L}^{\prime+}}Q^{\prime}c\left(1\right)\left|I\right\rangle \,,
\label{eq:energyprime}
\end{eqnarray}
Eqs. (\ref{eq:Qb})(\ref{eq:Qc}) are also rewritten as
\begin{eqnarray}
e^{-\epsilon\mathcal{L}^{\prime+}}Q^{\prime}b\left(1\right)\left|I\right\rangle  & = & e^{-\epsilon\mathcal{L}^{\prime+}}\left|I\right\rangle \,,\label{eq:Qb2}\\
e^{-\epsilon\mathcal{L}^{\prime+}}Q^{\prime}c\left(1\right)\left|I\right\rangle  & = & 0\,.\label{eq:Qc2}
\end{eqnarray}

Let us check if one can prove (\ref{eq:Qb2})(\ref{eq:Qc2}) by using
the expression (\ref{eq:Qprime3}). Substituting (\ref{eq:Qprime3})
into the left hand side of (\ref{eq:Qb2}), we get
\[
-\frac{1}{4}e^{-\epsilon\mathcal{L}^{\prime+}}UQU^{-1}b\left(1\right)\left|I\right\rangle \,.
\]
In order to avoid the singularity which appears in moving the operator
$U^{-1}$ to the right, we shift the position of $b$ for regularization.
Thus we consider 
\begin{eqnarray}
 &  & -\frac{1}{4}\lim_{\xi\to1}e^{-\epsilon\mathcal{L}^{\prime+}}UQU^{-1}b\left(\xi\right)\left|I\right\rangle \nonumber \\
 &  & \quad=-\frac{1}{4}\lim_{\xi\to1}\left[\frac{\left(\xi^{2}-1\right)^{2}}{\xi^{2}}Ue^{-\epsilon\tilde{\mathcal{L}}^{\prime+}}Qb\left(\xi\right)2\partial cc\left(1\right)\left|I\right\rangle \right]\,.\label{eq:Qb3}
\end{eqnarray}
where
\begin{eqnarray}
\tilde{\mathcal{L}}^{\prime+} & = & U^{-1}\mathcal{L}^{\prime+}U\nonumber \\
 & = & \left\{ Q,\,\frac{\pi}{2}\oint\frac{d\xi}{2\pi i}\left(1+\xi^{2}\right)\epsilon\left(\mathrm{Re}\xi\right)e^{h_{-\frac{1}{2}}\left(\xi\right)}b\left(\xi\right)\right\} \nonumber \\
 & = & \frac{\pi}{2}\oint\frac{d\xi}{2\pi i}\left(1+\xi^{2}\right)\epsilon\left(\mathrm{Re}\xi\right)e^{h_{-\frac{1}{2}}\left(\xi\right)}T\left(\xi\right)\,,\label{eq:Ltildeprime+}
\end{eqnarray}
Instead of $K^{\prime}$ or $\mathcal{L}^{\prime+}$, $\tilde{\mathcal{L}}^{\prime+}$
is the fundamental translation operator to deal with in the subsequent
calculation. Contrary to $K^{\prime}$, $\tilde{\mathcal{L}}^{\prime+}$
is made from $T\left(\xi\right)$ and we do not have to worry about
the conformal anomaly. If the operator $e^{-\epsilon\tilde{\mathcal{L}}^{\prime+}}$
should generate worldsheet around $\left\{ Q,b\left(\xi\right)\right\} $
in (\ref{eq:Qb3}), we could express $Q$ by a contour integral and
proceed further. 

\begin{figure}
\begin{centering}
\includegraphics[scale=0.5]{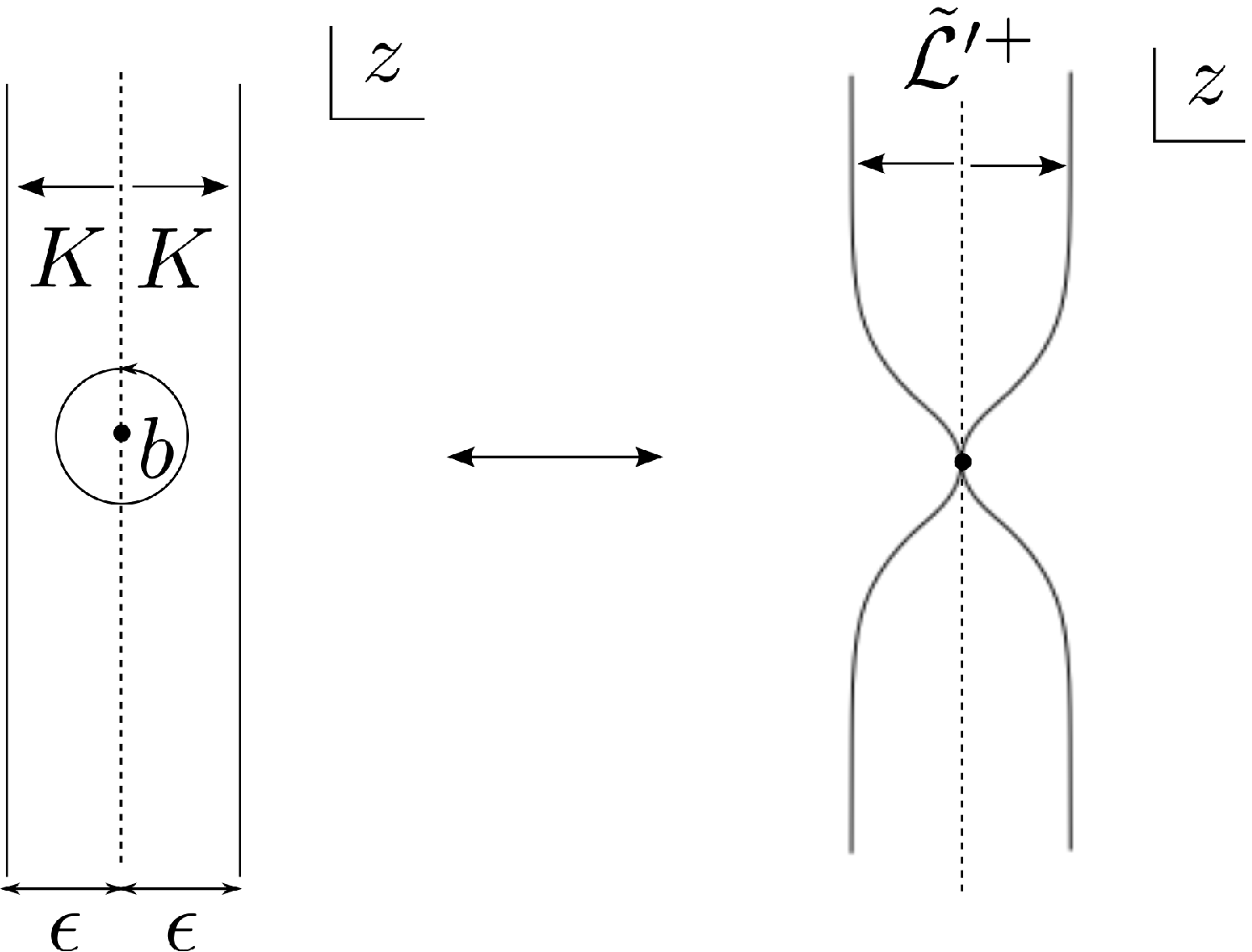}
\par\end{centering}

\caption{The worldsheet generated by $e^{-\epsilon\tilde{\mathcal{L}}^{\prime+}}$
in contrast to the one generated by $e^{-\epsilon K}$.\label{fig:The-worldsheet-generated}}
\end{figure}

The operator of the form (\ref{eq:Ltildeprime+}) can be analyzed
by the methods explained in \cite{Kiermaier2008}. Here it is convenient
to go to the sliver frame and rewrite (\ref{eq:Ltildeprime+}) as
\[
\tilde{\mathcal{L}}^{\prime+}=\int_{-i\infty}^{i\infty}\frac{dz}{2\pi i}e^{h\left(\frac{1}{2}+z\right)}T\left(\frac{1}{2}+z\right)+\int_{-i\infty}^{i\infty}\frac{dz}{2\pi i}e^{h\left(-\frac{1}{2}+z\right)}T\left(-\frac{1}{2}+z\right)\,,
\]
where 
\[
e^{h\left(z\right)}=-\frac{\cos^{2}\pi z}{\sin^{2}\pi z}\,.
\]
We introduce a new coordinate $w$ such that 
\[
\frac{\partial z}{\partial w}=e^{h\left(z\right)}\,,
\]
which is integrated as
\begin{equation}
w\left(z\right)=z-\frac{1}{\pi}\frac{\sin\pi z}{\cos\pi z}\,.\label{eq:wpm}
\end{equation}
Using these, $\tilde{\mathcal{L}}^{\prime+}$ is expressed as 
\[
\tilde{\mathcal{L}}^{\prime+}=\left[\int_{\frac{1}{2}-i\infty}^{\frac{1}{2}+i\infty}+\int_{-\frac{1}{2}-i\infty}^{-\frac{1}{2}+i\infty}\right]\frac{dz}{2\pi i}\frac{\partial w}{\partial z}T\left(w\right)\,,
\]
and $\tilde{\mathcal{L}}^{\prime+}$ generates translations with respect
to the coordinate $w$. The map $w\left(z\right)$ (\ref{eq:wpm})
maps the region $0<\mathrm{Im}z<\infty$ to $-\infty<\mathrm{Im}w<\infty$
and the region $-\infty<\mathrm{Im}z<0$ to $-\infty<\mathrm{Im}w<\infty$
for $\mathrm{Re}z=\pm\frac{1}{2}$ and $z=\pm\frac{1}{2}$ are singular
points. $z=\pm\frac{1}{2}$ are mapped to $w=\pm\infty$ and do not
move under the translation generated by $\tilde{\mathcal{L}}^{\prime+}$.
Therefore the operator $e^{-\epsilon\mathcal{L}^{\prime+}}$ acting
on the identity state $\left|I\right\rangle $ generates the worldsheet
of the form depicted in Fig. \ref{fig:The-worldsheet-generated}.
Hence $e^{-\epsilon\mathcal{L}^{\prime+}}$ in (\ref{eq:Qb3}) does
not generate worldsheet around $Qb\left(\xi\right)$ and we cannot
proceed from (\ref{eq:Qb3}). The correlation functions which appear
on the right hand sides of eqs.(\ref{eq:gioprime})(\ref{eq:energyprime})
correspond to cylinders of the form $w\sim w+L$. Such a cylinder
is mapped to two spheres whose coordinates are given by $e^{\frac{2\pi i}{L}w}$. 

\subsection*{Regularization}

\begin{figure}
\begin{centering}
\includegraphics[scale=0.5]{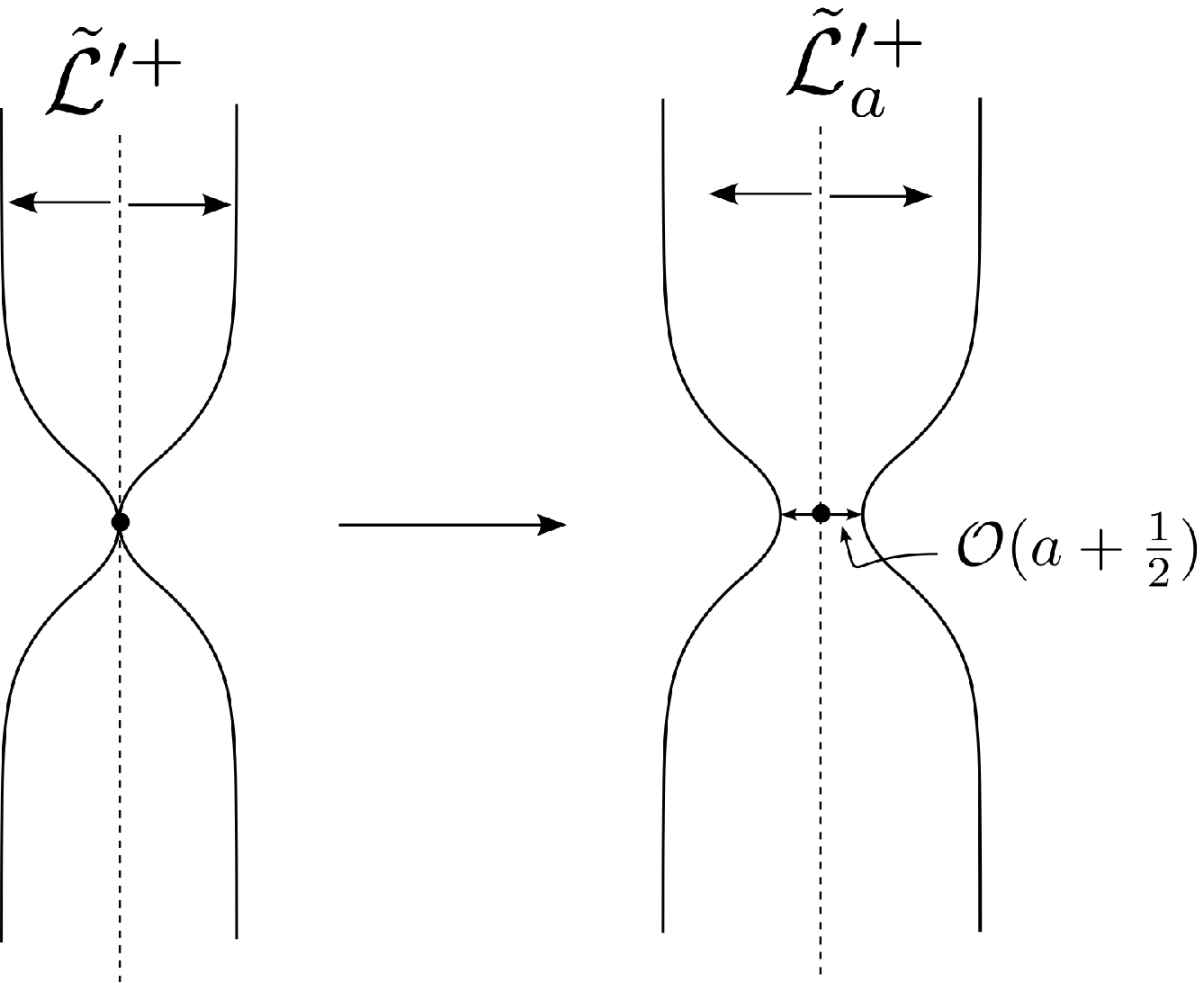}
\par\end{centering}

\caption{The surface generated by$e^{-L\tilde{\mathcal{L}}_{a}^{\prime+}}$
in contrast to the one generated by $e^{-L\tilde{\mathcal{L}}^{\prime+}}$.\label{fig:The-surface-generated}}
\end{figure}

The operator $e^{-\epsilon\tilde{\mathcal{L}}^{\prime+}}$ generates
apparently singular surfaces, which should be defined as a limit of
regular surfaces. There are problems in performing calculations on
such singular surfaces. We are not able to prove the homotopy relation
(\ref{eq:Qc2}) on such surfaces because no worldsheet is generated
around the point on the boundary. 
We would like to define the string field theory so that it describes
the tachyon vacuum. Therefore what we need to do is to regularize
the $\tilde{\mathcal{L}}^{\prime+}$, while preserving the relations
(\ref{eq:Qb2})(\ref{eq:Qc2}). The regularization we propose is to
replace $\tilde{\mathcal{L}}^{\prime+}$ by
\begin{equation}
\tilde{\mathcal{L}}_{a}^{\prime+}\equiv\frac{\pi}{2}\oint\frac{d\xi}{2\pi i}\left(1+\xi^{2}\right)\epsilon\left(\mathrm{Re}\xi\right)e^{h_{a}\left(\xi\right)}T\left(\xi\right)\,.\label{eq:L+a}
\end{equation}
 $\left(a>-\frac{1}{2}\right)$ with $h_{a}\left(\xi\right)$ given
in (\ref{eq:hxi}). We define $e^{-L\tilde{\mathcal{L}}^{\prime+}}$
as
\begin{equation}
\lim_{a\to-\frac{1}{2}}e^{-L\tilde{\mathcal{L}}_{a}^{\prime+}}\,.\label{eq:regularization}
\end{equation}
For $a>-\frac{1}{2}$, the surface generated by $e^{-L\tilde{\mathcal{L}}_{a}^{\prime+}}$
is of the form depicted in Fig. \ref{fig:The-surface-generated} and
we realize $e^{-L\tilde{\mathcal{L}}^{\prime+}}$ as a singular limit
of $e^{-L\tilde{\mathcal{L}_{a}}^{\prime+}}$. 

With such a regularization, the right hand side of (\ref{eq:Qb3})
becomes 
\begin{eqnarray*}
 &  & -\frac{1}{4}\lim_{a\to-\frac{1}{2}}\lim_{\xi\to1}\left[\frac{\left(\xi^{2}-1\right)^{2}}{\xi^{2}}Ue^{-\epsilon\tilde{\mathcal{L}}_{a}^{\prime+}}Qb\left(\xi\right)2\partial cc\left(1\right)\left|I\right\rangle \right]\\
 &  & \quad=U\lim_{a\to-\frac{1}{2}}e^{-\epsilon\tilde{\mathcal{L}}_{a}^{\prime+}}2\partial cc\left(1\right)\left|I\right\rangle \,,
\end{eqnarray*}
which can be rewritten as
\begin{eqnarray}
U\lim_{a\to-\frac{1}{2}}e^{-\epsilon\tilde{\mathcal{L}}_{a}^{\prime+}}2\partial cc\left(1\right)\left|I\right\rangle  & = & U\lim_{a\to-\frac{1}{2}}e^{-\epsilon\tilde{\mathcal{L}}_{a}^{\prime+}}U^{-1}\left|I\right\rangle \nonumber \\
 & = & e^{-\epsilon\mathcal{L}^{\prime+}}\left|I\right\rangle \,,\label{eq:Qb5}
\end{eqnarray}
and we eventually get (\ref{eq:Qb2}). (\ref{eq:Qc2}) can be proved
in the same way:
\begin{eqnarray}
 &  & e^{-\epsilon\mathcal{L}^{\prime+}}Q^{\prime}c\left(1\right)\left|I\right\rangle \nonumber \\
 &  & \quad=-\frac{1}{4}\lim_{\xi\to1}e^{-\epsilon\mathcal{L}^{\prime+}}UQU^{-1}c\left(\xi\right)\left|I\right\rangle \nonumber \\
 &  & \quad=-\frac{1}{4}\lim_{a\to-\frac{1}{2}}\lim_{\xi\to1}\left[\frac{\xi^{2}}{\left(\xi^{2}-1\right)^{2}}Ue^{-\epsilon\tilde{\mathcal{L}}_{a}^{\prime+}}Qc\left(\xi\right)2\partial cc\left(1\right)\left|I\right\rangle \right]\nonumber \\
 &  & \quad=0\,.\label{eq:Qc3}
\end{eqnarray}
One can immediately show that the terms on the right hand side of
(\ref{eq:energyprime}) vanish by using (\ref{eq:Qc3}). In order
to show that the right hand side of (\ref{eq:gioprime}) vanishes,
we use (\ref{eq:Qb5}) to get
\begin{eqnarray}
 &  & \left\langle I\right|e^{-\frac{L}{4}\mathcal{L}^{\prime+}}c\left(1\right)V\left(i,-i,\right)e^{-\frac{L}{4}\mathcal{L}^{\prime+}}\left|I\right\rangle 
\nonumber\\
 &  & \quad=\lim_{a\to-\frac{1}{2}}\lim_{\xi\to1}\left\langle I\right|e^{-\frac{L}{4}\mathcal{L}^{\prime+}}c\left(\xi\right)V\left(i,-i,\right)Ue^{-\frac{L}{4}\tilde{\mathcal{L}}_{a}^{\prime+}}2\partial cc\left(1\right)\left|I\right\rangle 
\nonumber\\
 &  & \quad=\lim_{a\to-\frac{1}{2}}\lim_{\xi\to1}\left\langle I\right|Ue^{-\frac{L}{4}\mathcal{\tilde{\mathcal{L}}}_{a}^{\prime+}}\frac{\left(\xi^{2}-1\right)^{2}}{\xi^{2}}c\left(\xi\right)V\left(i,-i,\right)e^{-\frac{L}{4}\tilde{\mathcal{L}}_{a}^{\prime+}}2\partial cc\left(1\right)\left|I\right\rangle 
\nonumber\\
 &  & \quad=0\,.
\end{eqnarray}
Here, with the regularization, $\left\langle I\right|U$ is away from
the other operators $c\left(\xi\right),\partial cc\left(1\right)$
and it can be set to zero. 
Thus we have shown how to regularize and define the operator
$e^{-L\tilde{\mathcal{L}}^{\prime+}}$ so that we can derive (\ref{eq:Qb2})(\ref{eq:Qc2})(\ref{eq:gioTT})(\ref{eq:energyTT}).
These formulas imply that the string field theory describes the tachyon
vacuum.

\section{Conclusions and discussions\label{sec:Conclusions-and-discussions}}

In this paper, we have evaluated the observables of the Takahashi-Tanimoto's
scalar solution (\ref{eq:TTsolution}) with $a=-\frac{1}{2}$, by
studying the Erler-Schnabl solution in the string field theory expanded
around it. The results are consistent with the claim that the solution
corresponds to the tachyon vacuum. In the calculations, the string
field $K^{\prime}$ or its worldsheet operator counterpart plays crucial
roles. In the latter half of this paper, we study the operator $K^{\prime}$
using the similarity transformation proposed by Kishimoto and Takahashi.
We discuss how we should treat it in order to be consistent with the
claim that the background is the tachyon vacuum. 

The relation (\ref{eq:Qprime3}) will be useful to evaluate various
other quantities in the string field theory expanded around $\Psi_{\mathrm{TT}}$.
Since the solution is supposed to describe the tachyon vacuum, we
expect all the amplitudes involving open string states to vanish. On
the other hand, we may be able to calculate closed string amplitudes
using the string field theory \cite{Gaiotto:2001ji,Drukker2003a,Drukker:2005hr}.
In order to do such calculations, we should take Siegel gauge for
example and construct the propagators. We will need some regularization
like (\ref{eq:regularization}) to define the propagator. We leave
it as a future problem.

The operator $U,U^{-1}$ in (\ref{eq:Qprime3}) should be related
to the boundary condition changing operators which play crucial roles
in \cite{Erler2014,Kiermaier:2010cf}. Suppose that we formally%
\footnote{Since $U,U^{-1}$ involve operators like $U_{2},U_{-2}$, we are not
so sure if we could do such a decomposition.%
} divide the operators $U,U^{-1}$ into the left and right piece $U_{L},U_{R},\left(U^{-1}\right)_{L},\left(U^{-1}\right)_{R}$
so that the operator $U,U^{-1}$ acts on a string field $A$ as
\begin{eqnarray*}
UA & = & U_{L}AU_{R}\,,\\
U^{-1}A & = & \left(U^{-1}\right)_{L}A\left(U^{-1}\right)_{R}\,.
\end{eqnarray*}
$U_{L},U_{R},\left(U^{-1}\right)_{L},\left(U^{-1}\right)_{R}$ may
be regarded as some kind of boundary condition changing operators
and the identities given in subsection \ref{subsec:UU-1}
imply the OPE's of them. 
It
would be inspiring to study the Takahashi-Tanimoto background from
the point of view of these operators.

\section*{Acknowledgments}

We are grateful to I.$\:$Kishimoto, C.$\:$Maccaferri, T. Masuda,
and T.$\:$Takahashi for sharing their ideas on this topic. We would
like to acknowledge T.$\;$Erler and Y.$\:$Okawa for useful comments.
We also would like to thank the organizers of the conference ``String
field theory and related aspects VI, SFT2014'', especially L. Bonora,
for hospitality. This work was supported in part by Grant-in-Aid for
Scientific Research (C) (25400242) from MEXT.

\appendix

\section{Maccaferri's method\label{sec:Maccaferri's-method}}

In a recent paper \cite{Maccaferri2014}, Maccaferri considered a
special case of Zeze map \cite{Kishimoto2007}, which maps an identity-based
solution to a regular solution. In the case of the Takahashi-Tanimoto
solution (\ref{eq:TTsolution}) with $a=-\frac{1}{2}$, one obtains
\begin{equation}
\Psi_{\mathrm{TT}}\to\Psi_{\mathrm{reg.}}\equiv\left(1+B\frac{1-F\left(K\right)}{K}\Psi_{\mathrm{TT}}\right)\left(Q+\Psi_{\mathrm{TT}}\right)\left(1+B\frac{1-F\left(K\right)}{K}\Psi_{\mathrm{TT}}\right)^{-1}\,.\label{eq:MZmap}
\end{equation}
The Zeze map (\ref{eq:MZmap}) is a gauge transformation and we can
get a regular solution gauge equivalent to $\Psi_{\mathrm{TT}}$ by
choosing $F\left(K\right)$ appropriately. A convenient choice is
$F\left(K\right)=\frac{1}{1+k}$ and we get
\begin{equation}
\Psi_{\mathrm{reg.}}=\frac{1}{1+K}\Psi_{\mathrm{TT}}\frac{1}{1+K^{\prime}}-Q\left(\frac{1}{1+K}\Psi_{\mathrm{TT}}\frac{1}{1+K^{\prime}}\right)\,,\label{eq:Psi}
\end{equation}
 which appears to be a regular solution. From the expression (\ref{eq:Psi}),
it is straightforward to calculate the energy and the Ellwood invariant
and one obtains \cite{Maccaferri2014}
\begin{eqnarray}
S\left[\Psi_{\mathrm{reg.}}\right] & = & -\frac{1}{6g^{2}}\mathrm{Tr}\left[\frac{1}{1+K}c\frac{1}{1+K}Qc\right]+\frac{1}{6g^{2}}\mathrm{Tr}\left[\frac{1}{1+K^{\prime}}c\frac{1}{1+K^{\prime}}Q^{\prime}c\right]\,,\nonumber \\
\mathrm{Tr}_{V}\Psi_{\mathrm{reg.}} & = & \mathrm{Tr}_{V}\left[\frac{1}{1+K}c\right]-\mathrm{Tr}_{V}\left[\frac{1}{1+K^{\prime}}c\right]\,.\label{eq:observablesPsi}
\end{eqnarray}
The right hand sides of eq.(\ref{eq:observablesPsi}) can be written
as 
\begin{eqnarray*}
S\left[\Psi_{\mathrm{reg.}}\right] & = & S\left[\Psi_{\mathrm{ES}}\right]-S\left[\Psi_{\mathrm{ES}}^{\prime}\right]\,,\\
\mathrm{Tr}_{V}\Psi_{\mathrm{reg.}} & = & \mathrm{Tr}_{V}\Psi_{\mathrm{ES}}-\mathrm{Tr}_{V}\Psi_{\mathrm{ES}}^{\prime}\,,
\end{eqnarray*}
where $\Psi_{\mathrm{ES}},\Psi_{\mathrm{ES}}^{\prime}$ are the Erler-Schnabl
solutions given in (\ref{eq:ES})(\ref{eq:ESprime}). Thus the observables
of $\Psi_{\mathrm{reg.}}$ are obtained from those of the Erler-Schnabl
solution $\Psi_{\mathrm{ES}}^{\prime}$. Using $S\left[\Psi_{\mathrm{ES}}^{\prime}\right]=\mathrm{Tr}_{V}\Psi_{\mathrm{ES}}^{\prime}=0$
derived in section \ref{sec:The-Erler-Schnabl-solution}, we can see
that the observables of $\Psi_{\mathrm{reg.}}$ coincide with those
of the tachyon vacuum solution $\Psi_{\mathrm{ES}}$.

\subsection*{Singularities}

Actually, the calculation of the observables above suffers from singularities
discussed by Maccaferri \cite{Maccaferri2014}. In calculating the
action, one typically encounters quantities of the form
\begin{eqnarray}
\left\langle c\left(z\right)c\partial c\left(0\right)\right\rangle _{C_{L}} & = & -\left(\frac{L}{\pi}\right)^{2}\sin^{2}\frac{\pi z}{L}\,,\label{eq:ccdcCL}
\end{eqnarray}
where $\left\langle \cdot\right\rangle _{C_{L}}$ denotes the correlation
function on a semi-infinite cylinder with circumference $L$. (\ref{eq:ccdcCL})
diverges in the limit $\mathrm{Im}z\to\pm\infty$ for small enough
$L>0$ or in the limit $L\to0$ with $\mathrm{Im}z\ne0$. Since the
Takahashi-Tanimoto solution (\ref{eq:TTsolution}) involves an integral
of the ghost $c$ up to $\mathrm{Im}z=\pm\infty$, we have trouble
in calculating the action%
\footnote{We do not encounter such divergences in the calculation of the Ellwood
invariant or the overlap of $\Psi_{\mathrm{reg.}}$ with Fock space
states. %
}. 

Therefore we need to find a good regularization to calculate the action%
\footnote{In \cite{Zeze2014a}, the author modifies the form of the solution
(\ref{eq:TTsolution}) as was presented in \cite{Maccaferri2014} and avoids
the singularity. %
}. In \cite{Maccaferri2014}, a solution with 
\begin{equation}
F\left(K\right)=F_{\epsilon}\left(K\right)=\frac{e^{-\epsilon K}}{1+\left(1-\epsilon\right)K}\,,\label{eq:Fepsilon}
\end{equation}
$\left(0\leq\epsilon\leq1\right)$ in (\ref{eq:MZmap}) is considered
as a regularization. Let $\Psi_{\epsilon}$ denote the $\Psi_{\mathrm{reg.}}$
with this choice of $F\left(K\right)$. It is easy to see that 
\[
\Psi_{\epsilon}=\frac{1}{1+K_{\epsilon}}\left(\Psi_{\mathrm{TT}}-\Psi_{\mathrm{TT}}B_{\epsilon}\frac{1}{1+K_{\epsilon}^{\prime}}\Psi_{\mathrm{TT}}\right)\,,
\]
 where 
\begin{eqnarray}
c_{\epsilon} & = & c\frac{KB}{G_{\epsilon}\left(K\right)}c\,,\nonumber \\
B_{\epsilon} & = & B\frac{G_{\epsilon}\left(K\right)}{K}\,,\nonumber \\
K_{\epsilon} & = & QB_{\epsilon}=G_{\epsilon}\left(K\right)\,,\nonumber \\
J_{\epsilon} & = & \left\{ B_{\epsilon},\Psi_{\mathrm{TT}}\right\}\,, \nonumber \\
K_{\epsilon}^{\prime} & = & K_{\epsilon}+J_{\epsilon}
\,,\label{eq:epsilon}
\end{eqnarray}
and
\[
\frac{1}{1+K_{\epsilon}}=\frac{1}{1+G_{\epsilon}\left(K\right)}=\frac{e^{-\epsilon K}}{1+\left(1-\epsilon\right)K}\,.
\]
The $\Psi_{\epsilon}$ consists of wedge states of width not smaller
than $\epsilon$ with operator insertions and we can avoid the above-mentioned
divergences taking $\epsilon>\frac{1}{2}$.

$K_{\epsilon},B_{\epsilon},c_{\epsilon}$ in (\ref{eq:epsilon}) satisfy
the $KBc$ algebra \cite{Erler2010,Masuda2012,Erler2012} and it is
straightforward to show that the observables for the solution $\Psi_{\epsilon}$
coincide with the shift in those of the modified Erler-Schnabl solutions
\cite{Maccaferri2014}
\begin{eqnarray}
\Psi_{\mathrm{ES},\epsilon} & = & \frac{1}{1+K_{\epsilon}}\left(c_{\epsilon}+Q\left(B_{\epsilon}c_{\epsilon}\right)\right)\,,\label{eq:ES0}\\
\Psi_{\mathrm{ES},\epsilon}^{\prime} & = & \frac{1}{1+K_{\epsilon}^{\prime}}\left(c_{\epsilon}+Q^{\prime}\left(B_{\epsilon}c_{\epsilon}\right)\right)\,,\label{eq:ESPhi}
\end{eqnarray}
namely
\begin{eqnarray}
S\left[\Psi_{\epsilon}\right] & = & S\left[\Psi_{\mathrm{ES},\epsilon}\right]-S\left[\Psi_{\mathrm{ES},\epsilon}^{\prime}\right]\nonumber \\
 & = & -\frac{1}{6g^{2}}\mathrm{Tr}\left[\frac{1}{1+K_{\epsilon}}c_{\epsilon}\frac{1}{1+K_{\epsilon}}Qc_{\epsilon}\right]+\frac{1}{6g^{2}}\mathrm{Tr}\left[\frac{1}{1+K_{\epsilon}^{\prime}}c_{\epsilon}\frac{1}{1+K_{\epsilon}^{\prime}}Q^{\prime}c_{\epsilon}\right]\,.\nonumber \\
\mathrm{Tr}_{V}\Psi_{\epsilon} & = & \mathrm{Tr}_{V}\Psi_{\mathrm{ES},\epsilon}-\mathrm{Tr}_{V}\Psi_{\mathrm{ES},\epsilon}^{\prime}\nonumber \\
 & = & \mathrm{Tr}_{V}\frac{1}{1+K_{\epsilon}}c_{\epsilon}-\mathrm{Tr}_{V}\frac{1}{1+K_{\epsilon}^{\prime}}c_{\epsilon}\,,\label{eq:observableepsilon}
\end{eqnarray}

Now we can use (\ref{eq:observableepsilon}) to calculate the observables.
As is pointed in \cite{Maccaferri2014}, although $\Psi_{\epsilon}$
itself may involve singularities for small $\epsilon$, $\Psi_{\mathrm{ES},\epsilon}^{\prime}$
is regular for all $0\leq\epsilon\leq1$. Moreover one can show 
\begin{eqnarray*}
\frac{\partial}{\partial\epsilon}\Psi_{\mathrm{ES},\epsilon} & = & Q\Lambda+\Psi_{\mathrm{ES},\epsilon}\Lambda-\Lambda\Psi_{\mathrm{ES},\epsilon}\,,\\
\frac{\partial}{\partial\epsilon}\Psi_{\mathrm{ES},\epsilon}^{\prime} & = & Q^{\prime}\Lambda^{\prime}+\Psi_{\mathrm{ES},\epsilon}^{\prime}\Lambda^{\prime}-\Lambda^{\prime}\Psi_{\mathrm{ES},\epsilon}^{\prime}\,,
\end{eqnarray*}
where
\begin{eqnarray*}
\Lambda & = & B_{\epsilon}\frac{1}{1+K_{\epsilon}}\frac{\partial}{\partial\epsilon}\Psi_{\mathrm{ES},\epsilon}\,,\\
\Lambda^{\prime} & = & B_{\epsilon}\frac{1}{1+K_{\epsilon}^{\prime}}\frac{\partial}{\partial\epsilon}\Psi_{\mathrm{ES},\epsilon}^{\prime}\,.
\end{eqnarray*}
Since the observables $\mathrm{Tr}_{V}\Psi_{\mathrm{ES},\epsilon},\mathrm{Tr}_{V}\Psi_{\mathrm{ES},\epsilon}^{\prime},S\left[\Psi_{\mathrm{ES},\epsilon}\right],S\left[\Psi_{\mathrm{ES},\epsilon}^{\prime}\right]$
are gauge invariant quantities, they are independent of $\epsilon$
provided the gauge parameters $\Lambda,\Lambda^{\prime}$ are regular
string fields. Thus we can evaluate them choosing $\epsilon$ for
which the calculation is easy. The most convenient choice is $\epsilon=0$
and we get 
\begin{eqnarray}
\mathrm{Tr}_{V}\Psi_{\epsilon} & = & \mathrm{Tr}_{V}\Psi_{\mathrm{ES}}-\mathrm{Tr}_{V}\Psi_{\mathrm{ES}}^{\prime}\,,\label{eq:gio0}\\
S\left[\Psi_{\epsilon}\right] & = & S\left[\Psi_{\mathrm{ES}}\right]-S\left[\Psi_{\mathrm{ES}}^{\prime}\right]\,.\label{eq:energy0}
\end{eqnarray}
From (\ref{eq:gioTT})(\ref{eq:energyTT}), we can see that the observables
of $\Psi_{\epsilon}$ coincide with those of the tachyon vacuum solution
$\Psi_{\mathrm{ES}}$. 

Thus, by using the Maccaferri's method, it is possible to construct
regular solutions gauge equivalent to $\Psi_{\mathrm{TT}}$, calculate
the observables of them and show that they coincide with those of
the tachyon vacuum. In a sense, this gives a more direct derivation
of the observables of the identity-based solutions than the one given
in section \ref{sec:The-Erler-Schnabl-solution}. On the other hand,
since the gauge transformation (\ref{eq:MZmap}) transforms an identity-based
solution into a regular solution, the transformation itself might
be somewhat singular. Therefore if the observables (\ref{eq:observablesPsi})
can be identified with those of $\Psi_{\mathrm{TT}}$ may be debatable. 

Before closing this appendix, one comment is in order. The string
field theory expanded around the Takahashi-Tanimoto solution possesses
a classical solution $-\Psi_{\mathrm{TT}}$ corresponding to the perturbative
vacuum. Although the solution itself is an identity-based solution,
one can construct a solution gauge equivalent to it
\[
-\frac{1}{1+K^{\prime}}\Psi_{\mathrm{TT}}\frac{1}{1+K}+Q^{\prime}\left(\frac{1}{1+K^{\prime}}\Psi_{\mathrm{TT}}\frac{1}{1+K}\right)\,,
\]
by Maccaferri's method. The observables can be calculated at least
formally and they coincide with those of the perturbative vacuum.

\section{Properties of $U,U^{-1}$\label{sec:Properties-of}}

In this appendix, we derive how the operators $U,U^{-1}$ act on the states $|0\rangle ,\langle 0|, |I\rangle ,\langle I|$. 

Let us first prove the following identities:
\begin{eqnarray}
U\left|0\right\rangle  
& = &
 \frac{1}{16}\partial bb\left(1\right)\partial bb\left(-1\right)c_{0}c_{1}\left|0\right\rangle \,,
\label{eq:U01}\\
U^{-1}\left|0\right\rangle  
& = & 
\frac{1}{16}\partial cc\left(1\right)\partial cc\left(-1\right)b_{-3}b_{-2}\left|0\right\rangle \,,
\label{eq:U02}\\
\left\langle 0\right|U 
& = & 
\left\langle 0\right|b_{2}b_{3}\,,
\label{eq:U03}\\
\left\langle 0\right|U^{-1} 
& = & 
\left\langle 0\right|c_{-1}c_{0}\,.
\label{eq:U04}
\end{eqnarray}
Since $q=2\sum_{n=1}^{\infty}\frac{1}{n}j_{-2n}$,
\[
e^{\pm q}\left|0\right\rangle =\exp\left[\pm2\sum_{n=1}^{\infty}\frac{1}{n}j_{-2n}\right]\left|0\right\rangle \,.
\]
 On the other hand, we have the bosonization formula 
\begin{eqnarray}
c\left(\xi\right) & = & \exp\left[\sum_{n=1}^{\infty}\frac{1}{n}j_{-n}\xi^{n}\right]e^{\sigma_{0}}e^{j_{0}\ln\xi}\exp\left[-\sum_{n=1}^{\infty}\frac{1}{n}j_{n}\xi^{-n}\right]\,,\label{eq:cbosonization}\\
b\left(\xi\right) & = & \exp\left[-\sum_{n=1}^{\infty}\frac{1}{n}j_{-n}\xi^{n}\right]e^{-\sigma_{0}}e^{-j_{0}\ln\xi}\exp\left[\sum_{n=1}^{\infty}\frac{1}{n}j_{n}\xi^{-n}\right]\,,\label{eq:bbosonization}
\end{eqnarray}
where $\sigma_{0}$ is the canonical conjugate of $j_{0}$ satisfying
\[
\left[j_{0},\sigma_{0}\right]=1\,.
\]
Eqs.(\ref{eq:cbosonization})(\ref{eq:bbosonization}) imply
\begin{eqnarray*}
\partial bb\left(1\right)\partial bb\left(-1\right)c_{-2}c_{-1}c_{0}c_{1}\left|0\right\rangle  & = & 16e^{-q}\left|0\right\rangle \,,\\
\partial cc\left(1\right)\partial cc\left(-1\right)b_{-5}b_{-4}b_{-3}b_{-2}\left|0\right\rangle  & = & 16e^{q}\left|0\right\rangle \,.
\end{eqnarray*}
From these, we get 
\begin{eqnarray*}
U\left|0\right\rangle  & = & e^{-q}U_{2}\left|0\right\rangle \\
 & = & e^{-q}\lim_{\varepsilon\to0}\partial bb\left(\varepsilon\right)\left|0\right\rangle \\
 & = & 
\lim_{\varepsilon\to0}
\left(
1-\frac{1}{\varepsilon^2}
\right)^{-4}
\partial bb\left(\varepsilon\right)\frac{1}{16}\partial bb\left(1\right)\partial bb\left(-1\right)c_{-2}c_{-1}c_{0}c_{1}\left|0\right\rangle \\
 & = & \frac{1}{16}\partial bb\left(1\right)\partial bb\left(-1\right)c_{0}c_{1}\left|0\right\rangle \,,\\
U^{-1}\left|0\right\rangle  & = & U_{-2}e^{q}\left|0\right\rangle \\
 & = & U_{-2}\frac{1}{16}\partial cc\left(1\right)\partial cc\left(-1\right)b_{-5}b_{-4}b_{-3}b_{-2}\left|0\right\rangle \\
 & = & \frac{1}{16}\partial cc\left(1\right)\partial cc\left(-1\right)b_{-3}b_{-2}\left|0\right\rangle \,.
\end{eqnarray*}
(\ref{eq:U03})(\ref{eq:U04}) are obtained from 
\[
\left\langle 0\right|e^{\pm q}=\left\langle 0\right|\,.
\]

Next, we examine how $U,U^{-1}$ act on $|I\rangle$. We will show
\begin{eqnarray}
U\left|I\right\rangle  
& = & 
\frac{1}{32}\partial bb\left(1\right)\left|I\right\rangle \,,
\label{eq:UI1}\\
U^{-1}\left|I\right\rangle  
& = & 
2\partial cc\left(1\right)\left|I\right\rangle \,.
\label{eq:UI2}
\end{eqnarray}
These
are shown by using the defining relation \cite{Rastelli:2000iu,Kishimoto:2001de,Schnabl:2002gg}
of $\left\langle I\right|$
\begin{equation}
\left\langle I\right|\phi\left(0\right)\left|0\right\rangle =\left\langle f\circ\phi\left(0\right)\right\rangle _{\mathrm{UHP}}\,,\label{eq:Idef}
\end{equation}
where 
\[
f\left(\xi\right)=\frac{2\xi}{1-\xi^{2}}\,,
\]
and $\left\langle \cdot\right\rangle _{\mathrm{UHP}}$ denotes the
correlation function on the upper half plane. In order to derive (\ref{eq:UI2}), for example, 
what we should do is to calculate
\[
\left\langle 0\right|\phi\left(0\right)U\left|I\right\rangle 
\,,
\]
and show that it is equal to $\left\langle 0\right|\phi\left(0\right)2\partial cc(1)|I\rangle$ 
for any $\phi\left(0\right)$. Since $U$
only changes the ghost part of $\left\langle I\right|$, 
we only have to deal with the case where $\phi\left(0\right)$ is made from 
ghost operators. 
Therefore what we
should calculate are the quantities of the form 
\begin{equation}
\left\langle 0\right|
\prod_{i}
c\left(\xi_{i}\right)
\prod_{j}
b\left(\xi_{j}^{\prime}\right)
U\left|I\right\rangle \,.
\label{eq:Ubccontour}
\end{equation}
Using eqs.(\ref{eq:U04}), (\ref{eq:UcU-1}), (\ref{eq:UbU-1}), we obtain
\begin{eqnarray}
 &  & \left\langle 0\right|\prod_{i}c\left(\xi_{i}\right)\prod_{j}b\left(\xi_{j}^{\prime}\right)U^{-1}\left|I\right\rangle 
\nonumber\\
 &  & \quad=\left\langle 0\right|c_{-1}c_{0}\prod_{i}\left(\frac{\left(\xi_{i}^{2}-1\right)^{2}}{\xi_{i}^{2}}c\left(\xi_{i}\right)\right)\prod_{j}\left(\frac{\xi_{j}^{\prime2}}{\left(\xi_{j}^{\prime2}-1\right)^{2}}b\left(\xi_{j}^{\prime}\right)\right)\left|I\right\rangle 
\nonumber\\
 &  & \quad=\left\langle I\right|\prod_{i}\left(\frac{\left(\xi_{i}^{2}-1\right)^{2}}{\xi_{i}^{2}}I\circ c\left(\xi_{i}\right)\right)\prod_{j}\left(\frac{\xi_{j}^{\prime2}}{\left(\xi_{j}^{\prime2}-1\right)^{2}}I\circ b\left(\xi_{j}^{\prime}\right)\right)c_{0}c_{1}\left|0\right\rangle 
\nonumber\\
 &  & \quad=\left\langle 0\right|\prod_{i}\left(\left(\frac{2}{f\left(\xi_{i}\right)}\right)^{2}f\circ I\circ c\left(\xi_{i}\right)\right)\prod_{j}\left(\left(\frac{f\left(\xi_{j}^{\prime}\right)}{2}\right)^{2}f\circ I\circ b\left(\xi_{j}^{\prime}\right)\right)\frac{1}{2}c_{0}c_{1}\left|0\right\rangle 
\nonumber\\
 &  & \quad=2\left\langle 0\right|U_{-2}\prod_{i}f\circ I\circ c\left(\xi_{i}\right)\prod_{j}f\circ I\circ b\left(\xi_{j}^{\prime}\right)\left|0\right\rangle 
\nonumber\\
 &  & \quad=2\left\langle 0\right|c_{-1}c_{0}\prod_{i}f\circ I\circ c\left(\xi_{i}\right)\prod_{j}f\circ I\circ b\left(\xi_{j}^{\prime}\right)\left|0\right\rangle
\nonumber \\
 &  & \quad=2\left\langle 0\right|f\circ I\circ\left(\partial cc\right)\left(1\right)\prod_{i}f\circ I\circ c\left(\xi_{i}\right)\prod_{j}f\circ I\circ b\left(\xi_{j}^{\prime}\right)\left|0\right\rangle 
\nonumber\\
 &  & \quad=\left\langle 0\right|\prod_{i}c\left(\xi_{i}\right)\prod_{j}b\left(\xi_{j}^{\prime}\right)2\partial cc\left(1\right)\left|I\right\rangle 
\label{eq:U-1proof}
\,,
\end{eqnarray}
where 
\[
I:\ \xi\to-\frac{1}{\xi}\,,
\]
is the inversion map. 
Eq.(\ref{eq:U-1proof}) implies $U^{-1}\left|I\right\rangle = 
2\partial cc\left(1\right)\left|I\right\rangle $. 
Eq.(\ref{eq:UI1}) can be shown in the same way.

Although the state $|I\rangle $ is not included in the Fock space, 
the operators $U,U^{-1}$ are inverse to one another, when they are acting on it. 
Indeed,
\begin{eqnarray}
U\left( U^{-1}|I\rangle \right)
&=&
2U\partial cc\left(1\right)\left|I\right\rangle
\nonumber
\\ 
&=&
2U\lim_{\xi \to 1}\partial cc\left(\xi\right)\left|I\right\rangle
\nonumber
\\
&=&
2U\lim_{\xi \to 1}
\left(\frac{\left(\xi^{2}-1\right)^{2}}{\xi^{2}}\right)^2
\partial cc\left(\xi\right)
\frac{1}{32}\partial bb\left(1\right)\left|I\right\rangle \,,
\nonumber
\\
&=&
\left|I\right\rangle\,,
\end{eqnarray}
and we can also get $U^{-1}\left( U|I\rangle \right)=|I\rangle$ in the same way.

Now let us consider the action of $U,U^{-1}$ on $\langle I|$. 
In order to get $\langle I|U$, we need to calculate 
\begin{equation}
\langle I|U\prod_{i}c\left(\xi_{i}\right)\prod_{j}b\left(\xi_{j}^{\prime}\right)\left|0\right\rangle
\,.
\end{equation}
Using (\ref{eq:U-1cU})(\ref{eq:U-1bU})(\ref{eq:U01}), it is straightforward
to get
\begin{eqnarray*}
 &  & U\prod_{i}c\left(\xi_{i}\right)\prod_{j}b\left(\xi_{j}^{\prime}\right)\left|0\right\rangle \\
 &  & \quad=\frac{1}{16}\prod_{i}\left(\frac{\left(\xi_{i}^{2}-1\right)^{2}}{\xi_{i}^{2}}c\left(\xi_{i}\right)\right)\prod_{j}\left(\frac{\xi_{j}^{\prime2}}{\left(\xi_{j}^{\prime2}-1\right)^{2}}b\left(\xi_{j}^{\prime}\right)\right)\partial bb\left(1\right)\partial bb\left(-1\right)c_{0}c_{1}\left|0\right\rangle \,.
\end{eqnarray*}
Now using (\ref{eq:Idef}), we obtain
\begin{eqnarray}
 &  & \left\langle I\right|U\prod_{i}c\left(\xi_{i}\right)\prod_{j}b\left(\xi_{j}^{\prime}\right)\left|0\right\rangle \nonumber \\
 &  & \quad=\left\langle 0\right|f\circ\left(\partial bb\right)\left(1\right)f\circ\left(\partial bb\right)\left(-1\right)\prod_{i}\left(\frac{\left(\xi_{i}^{2}-1\right)^{2}}{\xi_{i}^{2}}f\circ c\left(\xi_{i}\right)\right)\prod_{j}\left(\frac{\xi_{j}^{\prime2}}{\left(\xi_{j}^{\prime2}-1\right)^{2}}f\circ b\left(\xi_{j}^{\prime}\right)\right)f\circ\left(\partial cc\right)\left(0\right)\left|0\right\rangle \,.\nonumber \\
 &  & \ \label{eq:IUcb0}
\end{eqnarray}
Since 
\begin{eqnarray*}
f\circ\left(\partial bb\right)\left(\pm1\right) & = & \lim_{\varepsilon\to0}\left(\frac{\partial f}{\partial\xi}\right)^{5}\partial bb\left.\left(\frac{2\xi}{1-\xi^{2}}\right)\right|_{\xi=\pm1+\varepsilon}\\
 & = & \lim_{\varepsilon\to0}\varepsilon^{-10}\partial bb\left(-\frac{1}{\varepsilon}\right)\\
 & \sim & b_{2}b_{3}\,,
\end{eqnarray*}
acting on $\left\langle 0\right|$, 
\begin{equation}
\left\langle I\right|U\prod_{i}c\left(\xi_{i}\right)\prod_{j}b\left(\xi_{j}^{\prime}\right)\left|0\right\rangle
=
0
\,,
\end{equation}
provided
none of $\xi_{i},\xi_{j}^{\prime}$ coincides with $\pm1$. 
We can also derive, for example, 
\begin{equation}
\left\langle I\right|U
\partial cc(\pm 1)
\prod_{i}c\left(\xi_{i}\right)\prod_{j}b\left(\xi_{j}^{\prime}\right)\left|0\right\rangle
=
32
\left\langle I\right|
\prod_{i}c\left(\xi_{i}\right)\prod_{j}b\left(\xi_{j}^{\prime}\right)\left|0\right\rangle
\,,
\end{equation}
if none of $\xi_{i},\xi_{j}^{\prime}$ coincides with $\pm1$. 
Therefore we can set  $\langle I|U$ to zero in the case where there are no ghost operator insertions at $\xi =\pm 1$. 
One can show that $\langle I|U^{-1}$ can be set to zero in such situations, in the same way.  
However $\langle I|U\partial cc(\pm 1)$ and $\langle I|U^{-1}\partial bb(\pm 1)$ are not zero 
identically.  
We do not know how to express $\langle I|U$ and $\langle I|U^{-1}$ with such properties 
in a closed form. 

The vanishing of  $\langle I|U,\,\langle I|U^{-1}$ in some situations does not mean that the operators 
$U,U^{-1}$ are not invertible. 
For example, if one considers correlation function of the form 
\begin{equation}
\left(\left\langle I\right|U\right)
U^{-1}\prod_{i}c\left(\xi_{i}\right)\prod_{j}b\left(\xi_{j}^{\prime}\right)\left|0\right\rangle
\,,
\end{equation}
with $\xi_i\ne \pm 1, \xi_{j}^{\prime}\ne \pm 1$, one can see from  eq.(\ref{eq:U02}) that the operator 
$U^{-1}$ induces insertions of  $\partial cc(\pm 1)$:
\begin{eqnarray}
& &
\left(\left\langle I\right|U\right)
U^{-1}\prod_{i}c\left(\xi_{i}\right)\prod_{j}b\left(\xi_{j}^{\prime}\right)\left|0\right\rangle
\nonumber
\\
& &
\quad
=
\left\langle I\right|U
\frac{1}{16}\partial cc(1)\partial cc(-1)
\prod_i\left(\frac{\xi_i^2}{(\xi_i^2-1)^2}c(\xi_i)\right)
\prod_j\left(\frac{(\xi_j^{\prime 2}-1)^2}{\xi_j^{\prime 2}}b(\xi_j^\prime )\right)
b_{-3}b_{-2}|0\rangle
\,.
\end{eqnarray}
Hence we cannot set $\left\langle I\right|U$ to 
zero but rather we obtain
\begin{equation}
\left(\left\langle I\right|U\right)
U^{-1}\prod_{i}c\left(\xi_{i}\right)\prod_{j}b\left(\xi_{j}^{\prime}\right)\left|0\right\rangle
=
\left\langle I\right|\prod_{i}c\left(\xi_{i}\right)\prod_{j}b\left(\xi_{j}^{\prime}\right)\left|0\right\rangle
\,.
\end{equation}

\bibliographystyle{utphys}
\bibliography{identity6-6}

\end{document}